\documentclass[10pt,conference]{IEEEtran}
\usepackage{cite}
\usepackage{amsmath,amssymb,amsfonts}
\usepackage{algorithmic}
\usepackage{graphicx}
\usepackage{textcomp}
\usepackage{xcolor}
\usepackage[hyphens]{url}
\usepackage{fancyhdr}
\usepackage{hyperref}
\usepackage{listings}
\usepackage{color,soul}
\definecolor{codegreen}{rgb}{0,0.6,0}
\definecolor{codegray}{rgb}{0.5,0.5,0.5}
\definecolor{codepurple}{rgb}{0.58,0,0.82}
\definecolor{backcolour}{rgb}{0.95,0.95,0.92}
\lstdefinestyle{mystyle}{
    backgroundcolor=\color{backcolour},   
    commentstyle=\color{codegreen},
    keywordstyle=\color{magenta},
    numberstyle=\tiny\color{codegray},
    stringstyle=\color{codepurple},
    basicstyle=\ttfamily\footnotesize,
    breakatwhitespace=false,         
    breaklines=true,                 
    captionpos=b,                    
    keepspaces=true,                                  
    numbersep=5pt,                  
    showspaces=false,                
    showstringspaces=false,
    showtabs=false,                  
    tabsize=2
}
\newcommand{\hpcayear}{2025}

\newcommand{\hpcasubmissionnumber}{171}
\title{Reuse-Aware Compilation for Zoned Quantum Architectures Based on Neutral Atoms}

\def\hpcacameraready{} 

\newcommand\hpcaauthors{Wan-Hsuan Lin$^\dagger$, Daniel Bochen Tan$^{\dagger, \ddag}$, and Jason Cong$^\dagger$}
\newcommand\hpcaaffiliation{$^\dagger$University of California, Los Angeles, $^\ddag$Department of Physics, Harvard University}
\newcommand\hpcaemail{\{wanhsuanlin, bochentan\}@g.ucla.edu, cong@cs.ucla.edu}

\author{
  \ifdefined\hpcacameraready
    \IEEEauthorblockN{\hpcaauthors{}}
      \IEEEauthorblockA{
        \hpcaaffiliation{} \\
        \hpcaemail{}
      }
  \else
    \IEEEauthorblockN{\normalsize{HPCA \hpcayear{} Submission
      \textbf{\#\hpcasubmissionnumber{}}} \\
      \IEEEauthorblockA{
        Confidential Draft \\
        Do NOT Distribute!!
      }
    }
  \fi 
}

\fancypagestyle{camerareadyfirstpage}{%
  \fancyhead{}
  
  \fancyhead[C]{
    \ifdefined\aeopen
    \parbox[][12mm][t]{13.5cm}{\hpcayear{} IEEE International Symposium on High-Performance Computer Architecture (HPCA)}    
    \else
      \ifdefined\aereviewed
      \parbox[][12mm][t]{13.5cm}{\hpcayear{} IEEE International Symposium on High-Performance Computer Architecture (HPCA)}
      \else
      \ifdefined\aereproduced
      \parbox[][12mm][t]{13.5cm}{\hpcayear{} IEEE International Symposium on High-Performance Computer Architecture (HPCA)}
      \else
      \parbox[][0mm][t]{13.5cm}{\hpcayear{} IEEE International Symposium on High-Performance Computer Architecture (HPCA)}
    \fi 
    \fi 
    \fi 
    \ifdefined\aeopen 
      \includegraphics[width=12mm,height=12mm]{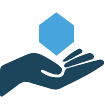}
    \fi 
    \ifdefined\aereviewed
      \includegraphics[width=12mm,height=12mm]{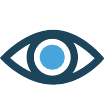}
    \fi 
    \ifdefined\aereproduced
      \includegraphics[width=12mm,height=12mm]{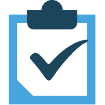}
    \fi
  }
  \fancyfoot[C]{}
}
\begin{document}
\maketitle

\ifdefined\hpcacameraready 
  \thispagestyle{camerareadyfirstpage}
  \pagestyle{empty}
\else
  \thispagestyle{plain}
  \pagestyle{plain}
\fi

\newcommand{\hpcaheight}{0mm}
\ifdefined\eaopen
\renewcommand{\hpcaheight}{12mm}
\fi

\begin{abstract}
Quantum computing architectures based on neutral atoms offer large scales and high-fidelity operations.
They can be heterogeneous, with different zones for storage, entangling operations, and readout.
Zoned architectures improve computation fidelity by shielding idling qubits in storage from side-effect noise, unlike monolithic architectures where all operations occur in a single zone.
However, supporting these flexible architectures with efficient compilation remains challenging.
In this paper, we propose ZAC, a scalable compiler for zoned architectures.
ZAC minimizes data movement overhead between zones with qubit reuse, i.e., keeping them in the entanglement zone if an immediate entangling operation is pending.
Other innovations include novel data placement and instruction scheduling strategies in ZAC, a flexible specification of zoned architectures, and an intermediate representation for zoned architectures, ZAIR.
Our evaluation shows that zoned architectures equipped with ZAC achieve a 22x improvement in fidelity compared to monolithic architectures.
Moreover, ZAC is shown to have a 10\% fidelity gap on average compared to the ideal solution.
This significant performance enhancement enables more efficient and reliable quantum circuit execution, enabling advancements in quantum algorithms and applications.
ZAC is open source at \href{https://github.com/UCLA-VAST/ZAC}{https://github.com/UCLA-VAST/ZAC}

\end{abstract}

\section{Introduction}

\begin{figure}[t]
    \centering
    \includegraphics[width=\linewidth]{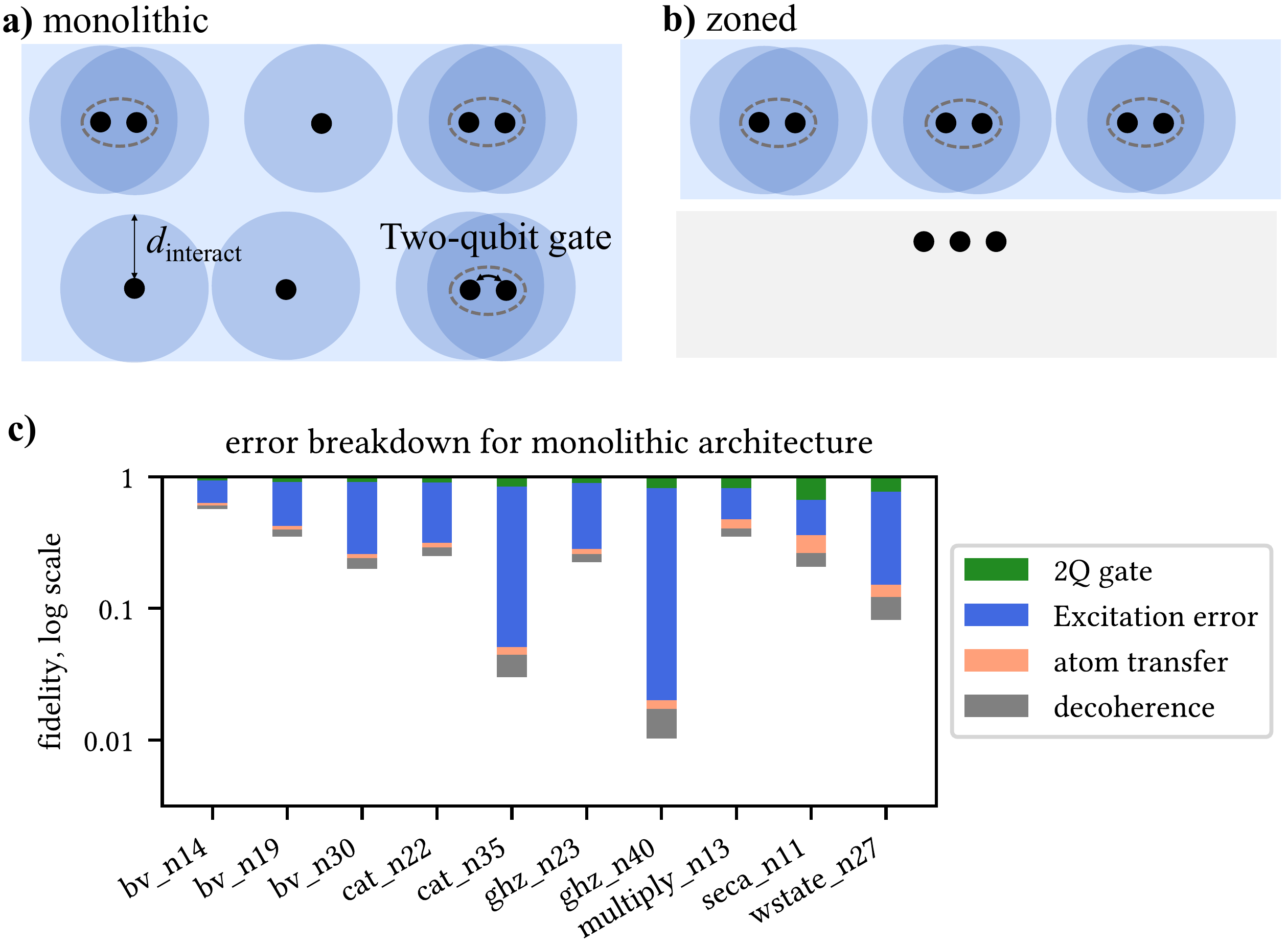}
    \caption{A comparison between \textbf{a)} monolithic and \textbf{b)} zoned architectures based on neutral atoms. 
    Blue regions are illuminated by the Rydberg laser.
    The grey region in b) is not covered by the Rydberg laser.
    \textbf{c)} Fidelity breakdown for the monolithic architecture based on the results in Ref.~\cite{tan2024enola}.
    Side-effect noise (blue, Rydberg excitation of idling qubits) is significant.}
    \label{fig:motivation}
\end{figure}

Rapid technological advances have established neutral atoms as a promising platform for quantum computing due to its scalability, long coherence time, and reconfigurability.
In these experiments, each atom needs to be held in a trap.
Using a spatial light modulator (SLM), large arrays of traps supporting thousands of qubits can be generated~\cite{pause2024supercharged, manetsch20246100tweezer, norcia2024ybatomarray}.
The key fidelity metric in quantum computing, two-qubit entangling gate fidelity, has reached 99.5\%\cite{evered2023high_fidelity_entanlging_gates}.
This gate is implemented by a laser that excites qubits to Rydberg states, illustrated as half-transparent discs in Fig.~\ref{fig:motivation}a.
When two qubits are within each other's discs (pairs in dashed ellipse), an entangling gate is performed.
Therefore, the placement of qubits in traps and the distance between the traps determine which qubit pairs can have entangling gates applied.
To change the placement, qubits can be transferred from SLM traps to mobile traps generated by an acousto-optic deflector (AOD) and moved accordingly.
This high-fidelity movement allows entangling gates to be executed on arbitrary pairs of qubits~\cite{bluvstein2022quantum}.

Prior compilation works~\cite{tan2022dpqa, tan2024dpqa, nottingham2023, wang2024qpilot, wang2024atomique, tan2024enola} primarily focused on the simplest, monolithic architecture, as shown in Fig.\ref{fig:motivation}a.
In this architecture, the Rydberg laser illuminates the entire region, exciting all qubits, including idle ones.
This side effect is a significant source of error.
In Fig.\ref{fig:motivation}c, we break down the fidelity estimation of running some quantum circuits on the monolithic architecture. 
These compiled results have the optimal number of Rydberg exposures, which means they have the minimum possible excitation errors.
However, these errors still dominate.
Zoned architectures, featuring distinct regions for entangling gates and qubit storage, have been demonstrated to address these errors~\cite{bluvstein2024logical}.
As illustrated in Fig.\ref{fig:motivation}b, the Rydberg laser is restricted to an entanglement zone (blue region), while a storage zone (grey region) shields idle qubits from Rydberg excitation. 

Preliminary works~\cite{decker2024arctic, stade2024abstract_model_for_zoned_arch} have begun exploring compilation for zoned architectures, but they do not fully leverage the benefits such zoned architectures offer.
Arctic~\cite{decker2024arctic} only supports one specific zoned architecture and introduce a large qubit movement overhead without qubit reuse.
NALAC~\cite{stade2024abstract_model_for_zoned_arch} aims at reducing movement overhead between zones via qubit reuse
but at the cost of increasing moving distance within the entanglement zone and exposing idle qubits to Rydberg excitation errors, lowering overall circuit fidelity. 

\color{black}
In this paper, we address compilation for zoned architectures based on neutral atoms aiming at supporting advanced zoned architectures and optimizing circuit performance via optimized qubit reuse.
Our key contributions are:
\begin{itemize}
    \item 
        We develop a novel \underline{z}oned \underline{a}rchitecture \underline{c}ompiler, named ZAC, consisting of a reuse-aware placement strategy and a load-balancing scheduling for architectures with multiple AODs and different zone configurations.
    \item 
        For placement, we design the cost function to approximate the movement duration which reduces qubit movement for both current and future circuit execution.
    \item
        We propose an optimized qubit reuse strategy as a key technique to reduce data movement overhead between entanglement and storage zones without increasing the excitation errors.
    \item 
        We present a load-balancing scheduling for rearrangement jobs, i.e., qubit movements, which minimizes execution time and maximizes hardware utilization by distributing the jobs across multiple AODs.
    \item 
        We support logical circuits with transversal gates, which is a fundamental operation in fault-tolerant quantum computing (FTQC). 
\end{itemize}

The evaluation results demonstrate a 22$\times$ fidelity improvement of a zoned architecture equipped with ZAC compared to monolithic architectures.
We also perform an optimality study showing that ZAC achieves near-optimal performance with only a 10\% fidelity gap from the ideal solution.

In addition to ZAC, we propose an architecture specification and intermediate representation (IR) for zoned architectures. 
The architecture specification enables precise definition and flexible configuration of multiple AODs, as well as storage and entanglement zones.
This facilitates the design of scalable and versatile architecture layouts.
Our \underline{z}oned \underline{a}rchitecture \underline{i}ntermediate \underline{r}epresentation, named ZAIR, strikes a balance between structure and detail by introducing the notion of rearrangement jobs.

The remainder of this paper is organized as follows.
We discuss related works in Section~\ref{sec:related_work} and presents our zoned architecture specification in Section~\ref{sec:spec_mz_arch}.
Section~\ref{sec:overview} gives the overview for our compiler, ZAC, 
and the detailed algorithms are provided in Section~\ref{sec:placement} and Section~\ref{sec:scheduling}.
Then, Section~\ref{sec:eval} presents the evaluation results.
Section~\ref{sec:ftqc} demonstrates applying ZAC in fault-tolerant quantum computing compilation.
Section~\ref{sec:ZAIR} introduces our proposed IR, ZAIR.
Finally, Section~\ref{sec:conclusion} concludes the paper and outlines future directions.

\section{Related Works}
\label{sec:related_work}
Early compilation works on quantum computing with neutral atoms (Baker et al.~\cite{baker2021exploiting}, Geyser~\cite{geyser}, TETRIS~\cite{ustc-neutral-atom}, and Schmid et al.~\cite{schmid2023}) assume individually addressable Rydberg entangling gates.
In this case, qubit routing leverages SWAP gates, similar to superconducting quantum processors, but these extra gates reduce circuit fidelity.
Moreover, individually addressed Rydberg gates need a different optical setup that has not yet achieved competitive fidelity (92.5\%~\cite{nature22-graham-atom-array}) compared to the 99.5\% fidelity of a global Rydberg laser~\cite{evered2023high_fidelity_entanlging_gates}).

OLSQ-DPQA~\cite{tan2022dpqa, tan2024dpqa} first targets ``dynamically field-programmable qubit array'' that is a monolithic architecture with a global Rydberg laser and atom movements.
However, it encodes the compilation to an SMT problem, which has exponential worst-case time complexity, limiting its scalability.
To address this, several scalable compilation ideas have emerged.
Q-Pilot~\cite{wang2024qpilot} leverages ``flying ancillas'' as an intermediate to perform entangling gates but suffers from many extra qubits and gates.
Nottingham et al.~\cite{nottingham2023} routes qubits with a ``switch'' operation that physically swaps the location of qubits, risking qubit collisions.
Atomique~\cite{wang2024atomique} takes a hybrid approach.
For inter-array entangling gates (with one qubit in the SLM and the other in the AOD), it moves the whole AOD array to get these pairs to interact.
For intra-array gates, SWAPs between the two arrays are inserted as necessary to make them inter-array gates later.
The latest work Enola~\cite{tan2024enola} employs graph theory in the compilation.
It first schedules the entangling gates with an edge-coloring algorithm for a near-optimal number of layers.
Then, it derives rounds of parallel qubit movements between the layers with a maximal independent set algorithm.

While the monolithic architecture's fidelity bottleneck, the global Rydberg exposure, has been optimized to near-optimal,
the noise on idle qubits is still significant (Fig.~\ref{fig:motivation}c), especially for deep circuits with many dependencies.
The zoned architecture can avoid this by moving only the necessary qubits to the entanglement zone.
Arctic~\cite{decker2024arctic} supports one specific zoned architecture: an entangling zone with a single row of traps ``sandwiched'' by two storage zones.
It demonstrates fidelity improvement over the monolithic architectures, but the applicability is limited.
NALAC~\cite{stade2024abstract_model_for_zoned_arch} supports zoned architectures with various sizes.
It moves two rows of qubits from the storage to the entangling zone each time and ``slides'' the two rows past each other to interact qubits,
which enables qubit reuse.
However, NALAC adopts a greedy placement strategy to minimize movement in one stage and limits gate placement to a single row in the entanglement zone, underutilizing the zone's capacity. 
Lastly, NALAC lacks support for architectures with multiple AODs and varied zone configurations, limiting its capability for architectural exploration and adaptation to advanced systems.
In contrast, ZAC can fully utilize the entanglement zone by placing gates across multiple rows and support compilation for advanced architectures.

\color{black}
\section{A General Specification of Zoned Architectures}
\label{sec:spec_mz_arch}

\begin{figure}
    \centering
    \includegraphics[width=\linewidth]{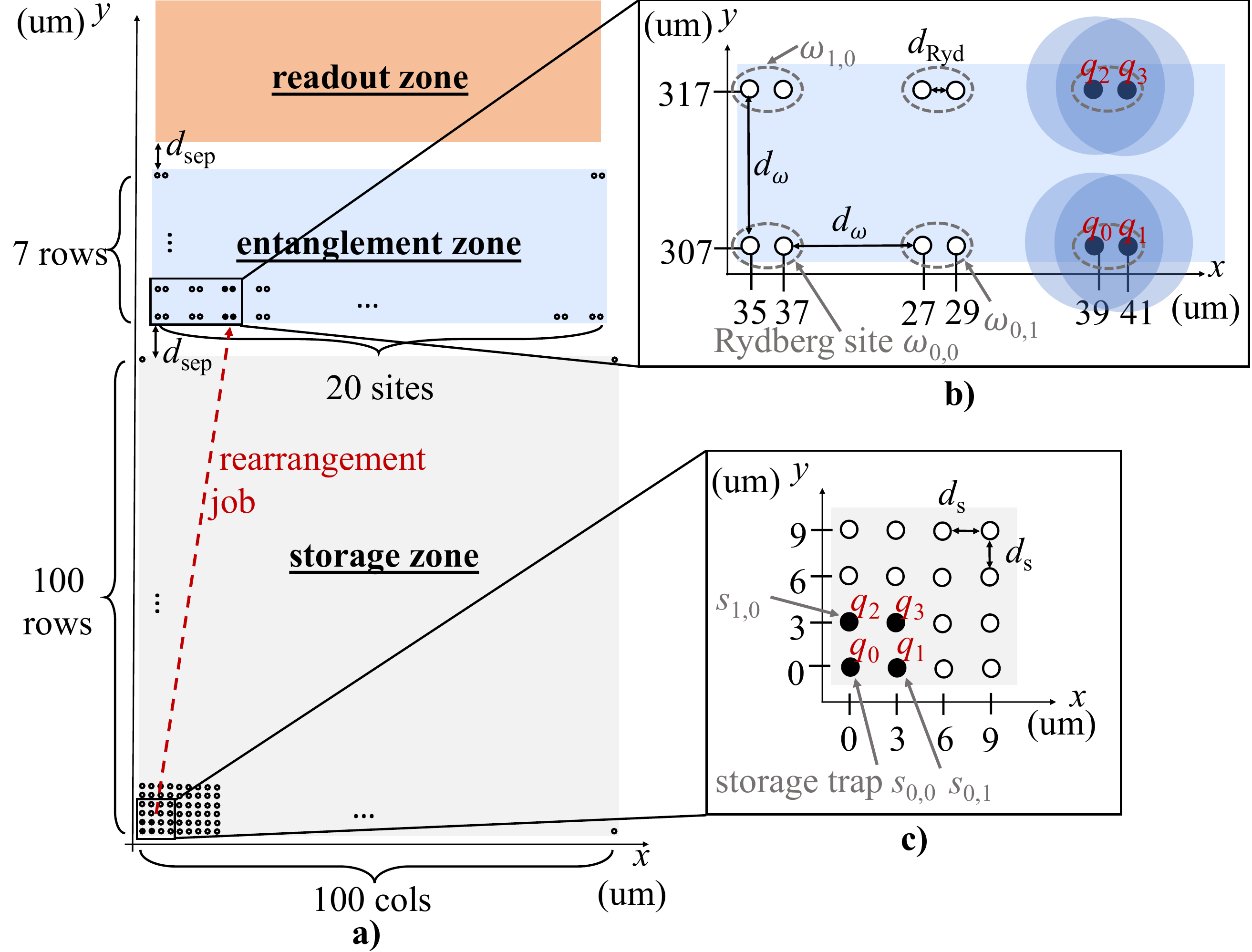}
    \caption{
Our reference zoned architecture following Ref.~\cite{bluvstein2024logical}.
\textbf{a)} Overview of the architecture consisting of a readout zone (orange) for qubit measurement, an entanglement zone (blue) for Rydberg entangling gates, and a storage zone (grey) for idle qubits.
The entanglement zone consists of 7 rows of Rydberg sites, each contains 20 sites.
The storage zone consists of 100$\times$100 storage traps.
The zone separation is $d_\text{sep}$=10um.
The red dashes denote a rearrangement job from the storage to the entanglement zone.
\textbf{b)} Detailed layout in the entanglement zone.
The dotted gray eclipses indicate Rydberg sites, each with two SLM traps separated by $d_\text{Ryd}$=2um.
The blue discs indicate the Rydberg range for qubit interaction.
The distance between two Rydberg sites is $d_\omega$=10um.
This prevent unwanted interaction between qubits in different sites.
The circles represent the empty traps, and the solid dots are the traps occupied by qubits. 
When the Rydberg laser is on, two qubits in the same Rydberg site perform a CZ operation,
e.g., CZ($q_0$,$q_1$) and CZ($q_2$,$q_3$).
\textbf{c)} Detailed layout in the storage zone, where the qubit separation is $d_s$=3um.}
    \label{fig:csarchitecture}
\end{figure}

\begin{figure}
    \centering
    \includegraphics[width=\linewidth]{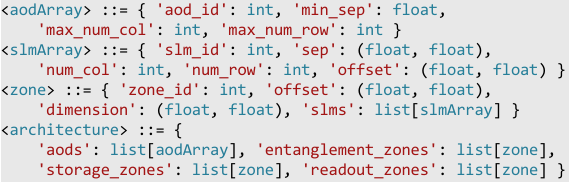}
    \caption{Specification of AOD arrays, SLM arrays, zones, and architecture. }
    \label{fig:spec}
\end{figure}

Our specification contains four types of entities: AOD arrays, SLM arrays, zones, and the architecture, as listed in Fig.~\ref{fig:spec}.
We shall explain the specification with our reference zoned architecture following Ref.~\cite{bluvstein2024logical} (Fig.~\ref{fig:csarchitecture}).

Each AOD array corresponds to an optical apparatus that is a product of a row component and a column component.
Each row and column can be turned on/off and moved during the circuit execution.
The AOD traps are the intersection points of the activated rows and columns.
We need an index \texttt{aod\_id} for each AOD array because there may be multiple AOD arrays in the architecture.
Each array will have a \texttt{max\_num\_col} and \texttt{max\_num\_row} reflecting the capacity of the column and row components.
The \texttt{min\_sep} is the minimum separation between any two rows and any two columns at any time.

An SLM array contains \texttt{num\_col} columns and \texttt{num\_row} rows of traps.
\texttt{sep} specifies the x and y separations.
To anchor the array at a specific location, we also need an \texttt{offset} which is the offset of the bottom left trap with respect to the origin in the x and y coordinate.
Each SLM array does not correspond to an individual optical apparatus.

This definition may seem a bit convoluted, but our goal is to support the configuration of the entanglement zones in a compact way.
In our reference architecture, the SLM traps in an entanglement zone (Fig.~\ref{fig:csarchitecture}b) need to form \textit{Rydberg sites} $\omega$.
These are pairs of traps separated by $d_\text{Ryd}$ = 2um.
The sites are indexed by row and column, e.g., $\omega_{1,0}$ is the Rydberg site at row 1 and column 0.
To prevent unwanted interactions between qubits in different Rydberg sites, the distance between two rows or columns of Rydberg sites is $d_\omega$ = 10um.
To describe the reference configuration, we specify two \texttt{slmArray}s with the same separation, but
with different offsets: (35, 307) and (37, 307).
As the qubits within a Rydberg site are placed in the same row, the x separation for the \texttt{slmArray}s is $d_\text{Ryd}+d_\omega$ = 12um, 
and the y separation is $d_\omega$ = 10um.

In the storage zone (Fig.~\ref{fig:csarchitecture}c), the SLM separations can be smaller, $d_\text{s}$ = 3um, because in this zone it is unnecessary to leave $d_\omega$ between qubits to avoid Rydberg interaction.
We denote $s_{r,c}$ as a storage trap at row $r$ and column $c$.

Each zone corresponds to a physical region with certain boundaries.
Thus, it has an \texttt{offset}, i.e., location of bottom left corner, and a \texttt{dimension} which are its width and height.
Inside a zone, there can be some SLM arrays.
In our reference architecture (Fig.~\ref{fig:csarchitecture}a), the readout zone has no SLM, the storage zone has one, and the entanglement zone has two.

Finally, the whole architecture is defined by a list of AOD arrays, a list of entanglement zones, a list of storage zones, and a list of readout zones.
In our reference architecture, there is only one zone of each type, and they are separated by $d_\text{sep}\text{ = 10um} \geq d_\omega$ to avoid any unwanted Rydberg interaction.

By providing this comprehensive and flexible specification, our architecture supports the customized and scalable designs, adaptable to a wide range of experimental setups and computational requirements.
For example, it is possible to increase the number of SLM traps in a Rydberg site to leverage a Rydberg gate on more qubits.
Another example is to specify multiple entanglement zones as we will demonstrate later.

\section{ZAC: Zoned Architecture Compiler}
\label{sec:overview}
\begin{figure}
    \centering
    \includegraphics[width=\linewidth]{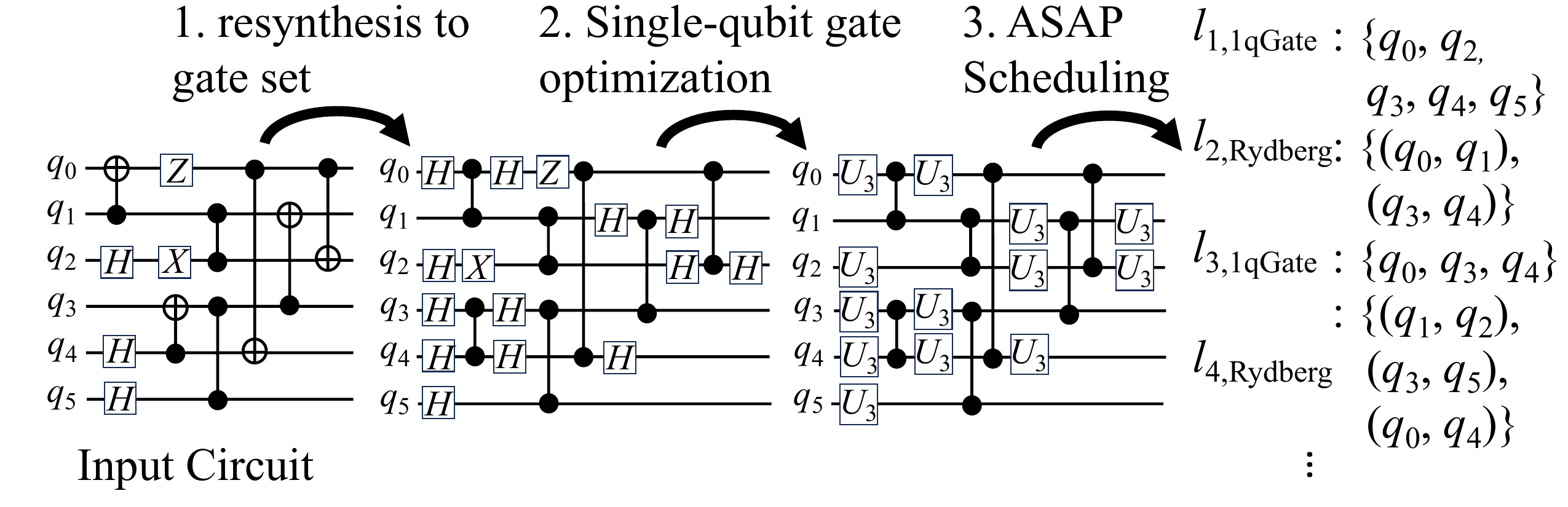}
    \caption{Preprocessing: resynthesis, 1Q gate optimization, and as-soon-as-possible 2Q gate scheduling. 
    The output is a list of gate stages.}
    \label{fig:preprocessing}
\end{figure}

Our compilation consists of three steps: preprocessing, placement, and scheduling.
During preprocessing, we resynthesize the circuit based on the hardware-supported gate set \{CZ, U3\} and optimize 1Q gates.
This can be done in standard approaches, e.g., with Qiskit~\cite{qiskit2024}.
Next, we assign gates into Rydberg and 1Q gate stages, 
and ensure that a qubit is involved in no more than one gate per stage.
Fig.~\ref{fig:preprocessing} provides an example of preprocessing.
This circuit is our running example in the following sections.

After preprocessing, the Rydberg stages can be readily translated into instructions in ZAIR.
However, the qubits are not yet mapped to traps.
In the placement step, we first derive the initial locations of qubits in the storage zone.
Then, for each Rydberg stage, we map each 2Q gate to a Rydberg site in an entanglement zone.
After each Rydberg stage, idle qubits for the next Rydberg stage need to be moved back to a storage zone to prevent Rydberg excitation errors.

Once placement is complete, we know the change of location for each qubit before and after each Rydberg stage.
In the scheduling step, we generate a set of \textit{rearrangement jobs} to accommodate these changes in qubit location, e.g., in Fig.~\ref{fig:csarchitecture} there is a job moving four qubits from the storage to the entanglement zone.
These jobs have dependencies with the other types of instructions and with each other.
Based on these dependencies, we schedule the instructions to optimize circuit execution time.
If there are multiple AODs, we assign the rearrangement jobs to AODs in a load-balanced manner.

With a zoned architecture, idle qubits can avoid Rydberg excitation in the storage zone, so the number of Rydberg stages is no longer the critical factor for fidelity.
Instead, the primary error sources are the overheads associated with qubit movements, as they incur errors from decoherence and atom transfers during qubit pickup and drop-off.
Since the changes in qubit locations determine which and how the qubits are moved, we find placement to be the most important step.

Our placement strategy includes two main innovations.
First, we utilize the concept of \textit{reuse}: a qubit in the entanglement zone is considered reusable if it will be involved in the next Rydberg stage.
These reusable qubits can remain at their Rydberg sites for the next gate operations, thus reducing atom transfer errors and qubit movement time.
Second, we refine the cost function for qubit placement.
Intuitively, the cost of a 2Q gate can be the distance between the two qubits in the gate.
As done in some previous works, one approach is minimizing the sum of all gate costs in the initial qubit placement.
However, this cost is not applicable in our case because the distance between two qubits can not directly reflect the movement distance for performing the gate at a Rydberg site.
Additionally, it does not consider movements that can be bundled together in the same rearrangement job.
Thus, we propose a cost function to better estimate the movement duration to a site and incorporate considerations for the parallel movements.
In the intermediate placements, we also enhance the cost function with a lookahead cost to minimize movement distances for future Rydberg stages.


\section{Reuse-Aware Placement}
\label{sec:placement}
\subsection{Initial Qubit Placement Based on Simulated Annealing}
\label{subsec:init_placement}
\begin{figure}
    \centering
    \includegraphics[width=\linewidth]{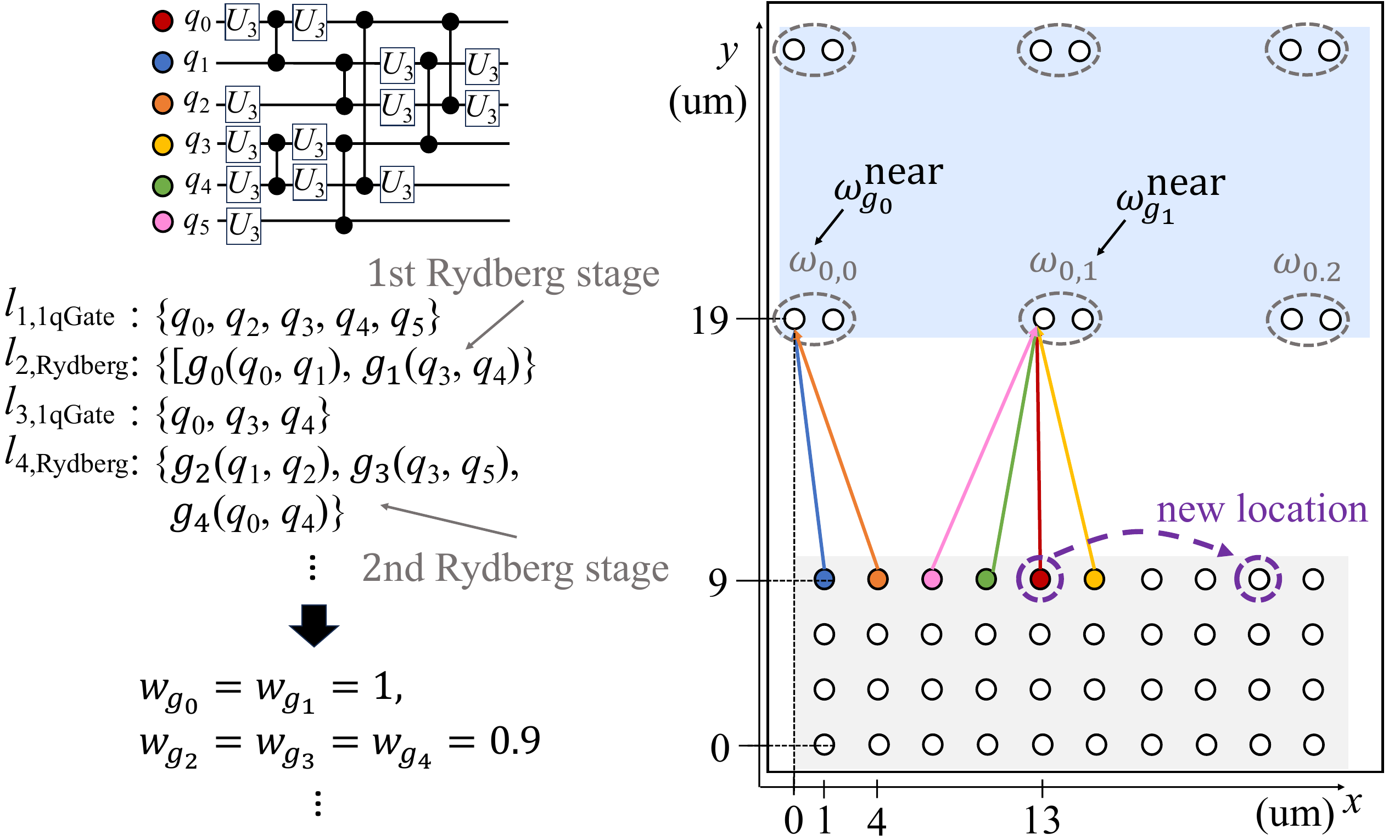}
    \caption{Initial placement.
    Qubits are represented by dots in different colors.
    Colored edges indicate the nearest Rydberg sites for each qubit, e.g., the nearest site for $q_0$ (red dot) is $\omega_{0,1}$.
    $\omega^\text{near}_{g_0}$ and $\omega^\text{near}_{g_1}$ are the nearest Rydberg site for $g_0$ and $g_1$, respectively.
    A possible movement in simulated annealing (dashed arrow) is to move $q_0$ to the site at $r_3$ and $c_9$ in the storage zone.
    }
    \label{fig:initial_placement}
\end{figure}

A good initial qubit placement aims to minimize movement overheads throughout the computation process. 
The movement overhead for a gate to a site is estimated based on the qubit distance to the site.
For the case where the movements can be performed in parallel, the cost is estimated by the \textit{maximum} of the distance between the qubits and the site because the movement duration is decided by the longest movement. 
Otherwise, the cost is estimated by the distance \textit{sum}. 
The movements can be performed in parallel if initially, the qubits are in the same SLM row.
Because they will be picked up by one AOD row, we can move them together to the Rydberg site via stretching the AOD,
e.g., in Fig.~\ref{fig:initial_placement}, $q_0$ and $q_1$ can be moved to site $\omega_{0,0}$ simultaneously.
However, if the qubits are not in the same row, different AOD rows will pick them up. 
As a result, they can not be dropped off at the Rydberg site simultaneously due to the non-stacking constraint for AOD rows.
Therefore, such movements are performed sequentially.

Formally, the qubit placement at a certain time are denoted by $M=(m_0, m_1, \ldots, m_n)$, where $m_q=(x_q,y_q)$ is the exact location for qubit $q$.
Then, given a qubit placement $M$, the movement cost for gate $g(q,q')$ to Rydberg site $\omega$ is
\begin{equation}
    \mathit{gCost}(g, \omega, M)=\begin{cases}
        \sqrt{\mathit{d}(\omega, m_q)} + \sqrt{\mathit{d}(\omega,m_{q'})},\text{ if } y_q \neq y_{q'}\\
        \max(\sqrt{\mathit{d}(\omega, m_q)}, \sqrt{\mathit{d}(\omega,m_{q'})}), o.w.    
    \end{cases} 
    \label{eq:est_gate_cost}
\end{equation}
A square root function is applied since the movement duration is proportional to the square root of movement distance~\cite{bluvstein2022quantum}.

The cost of the whole qubit placement, is estimated as the sum of the cost of all 2Q gates.
The cost for a gate $g$ is estimated based on its target qubits and their nearest Rydberg site $\omega^\text{near}_g$, designated as $\mathit{gCost}(g, \omega^\text{near}_g, M_0)$.
$\omega^\text{near}_g$ is determined by first identifying the nearest Rydberg sites for each target qubit, 
and then selecting the middle Rydberg site between these two sites, i.e., if $\omega_{r,c}$ and $\omega'_{r',c'}$ are the nearest Rydberg sites for target qubits $q$ and $q'$ respectively, then $\omega^\text{near}_g = \omega_{r^*,c^*}$, where $r^* = \lfloor (r+r')/2 \rfloor$ and $c^* = \lfloor (c+c')/2 \rfloor$.

Consider the example qubit placement shown in Fig.~\ref{fig:initial_placement}. 
For gate $g_0(q_0,q_1)$, the nearest Rydberg site is $\omega^\text{near}_{g_0}=\omega_{(0+0)/2,\lfloor(0+1)/2\rfloor}=\omega_{0,0}$.
Under placement on the right, $q_0$ is at $s_{3,4}$ with $x$ = 13 and $y$ = 9,
and $q_1$ is at $s_{3,0}$ with $x$ = 1 and $y$ = 9.
For distance calculations, we use the left trap in a Rydberg site as its reference location.
Thus, the exact location for $\omega_{0,0}$ is $x$ = 0 and $y$ = 19.
The distances are then computed as $\mathit{d}(\omega_{0,0}, s_{3,4})=\sqrt{(0-13)^2+(19-9)^2}=16.40$, 
and $\mathit{d}(\omega_{0,0}, s_{3,0})=10.05$.
Since $q_0$ and $q_1$ are in the same row, 
the cost for $g_0$ is $\max(\sqrt{16.40}, \sqrt{10.05})=4.05$.

Our cost for the whole initial placement $M$, is:
\begin{equation}
\label{eq:sa_cost}
    \mathit{cost}(M) = \sum_{g\in G_2} w_g\times \mathit{gCost}(g, \omega^{\text{near}}_g, M),
\end{equation}
where $G_2$ is the set of 2Q gates in the circuit.
We assign a weight factor $w_g$ for each gate, calculated as
$w_g = \max(0.1, 1 - 0.1(t - 1))$,
where $t$ is the Rydberg stage at which gate $g$ is scheduled to be executed.
For example, in Fig.~\ref{fig:initial_placement}, $w_{g_2}$ = 0.9, as $g_2$ belongs to the second Rydberg stage.
This weighting scheme prioritizes gates occurring earlier in the circuit.
The rationale behind this weighting is that with dynamic qubit placement, the locations of qubits in the storage zone change as the circuit execution progresses.
Consequently, using the initial qubit locations to estimate distances for gates later in the circuit may lead to inaccuracies.
By assigning higher weights to earlier gates, we ensure that the placement optimization focuses more on the initial stages of the circuit, where our distance estimates are most reliable.

We apply a simulated annealing (SA) framework~\cite{van1987sa} to minimize the cost.
During annealing, we generate neighboring states by changing the qubit location.
For example, a qubit may exchange locations with another qubit or jump to an empty trap, as indicated by the dashed arrow in Fig.~\ref{fig:initial_placement}.
The process terminates when the cost converges or the iteration exceeds a user-defined limit.
Based on our empirical results, we set a 1000 iteration limit.
The complexity for SA is $O(g)$, where $g$ is the number of 2Q gates, since we need to calculate a cost for each gate.
Although other placement techniques, e.g., analytical methods, may be applied as well, our experimental results (Section~\ref{sec:eval}) shows that our SA-based approach with maximal qubit reuse (will be presented in Section~\ref{subsec:intermediate_placement}) achieves near-optimal solutions.

\subsection{Reuse-Aware Dynamic Placement}
\label{subsec:intermediate_placement}
Since there are multiple Rydberg stages, the qubit placement can be dynamic.
Below we iteratively determine intermediate placements after the initial placement. 
If a qubit $q$ is reused in the next Rydberg stage by a gate $g(q,q')$, we would like to maintain it in the entanglement zone.
We choose to keep it at the same site.
Otherwise, $q$ still necessitates movements, albeit inside the same zone.
If a qubit is not reused, we assign it  a trap in the storage zone in a non-reuse dynamic qubit placement step.

During gate placement, we allocate Rydberg sites for the gates that do not reuse qubits.
While the qubit reuse idea can reduce movement of $q$, it may have a negative impact too, 
because $q'$ may be required to traverse longer distances to the current site of $q$.
Thus, for each Rydberg stage, we generate two placement solutions for non-reuse dynamic qubit placement and gate placement, one incorporating qubit reuse and the other without it.
We commit to the the better solution between the two.
Before the first Rydberg stage, no qubits are in entanglement zones, so we do not need to consider reuse.
The following subsections detail our approach to identify qubit reuse, place 2Q gates, and place non-reuse qubits back to the storage zone.

\begin{figure}
    \centering
    \includegraphics[width=\linewidth]{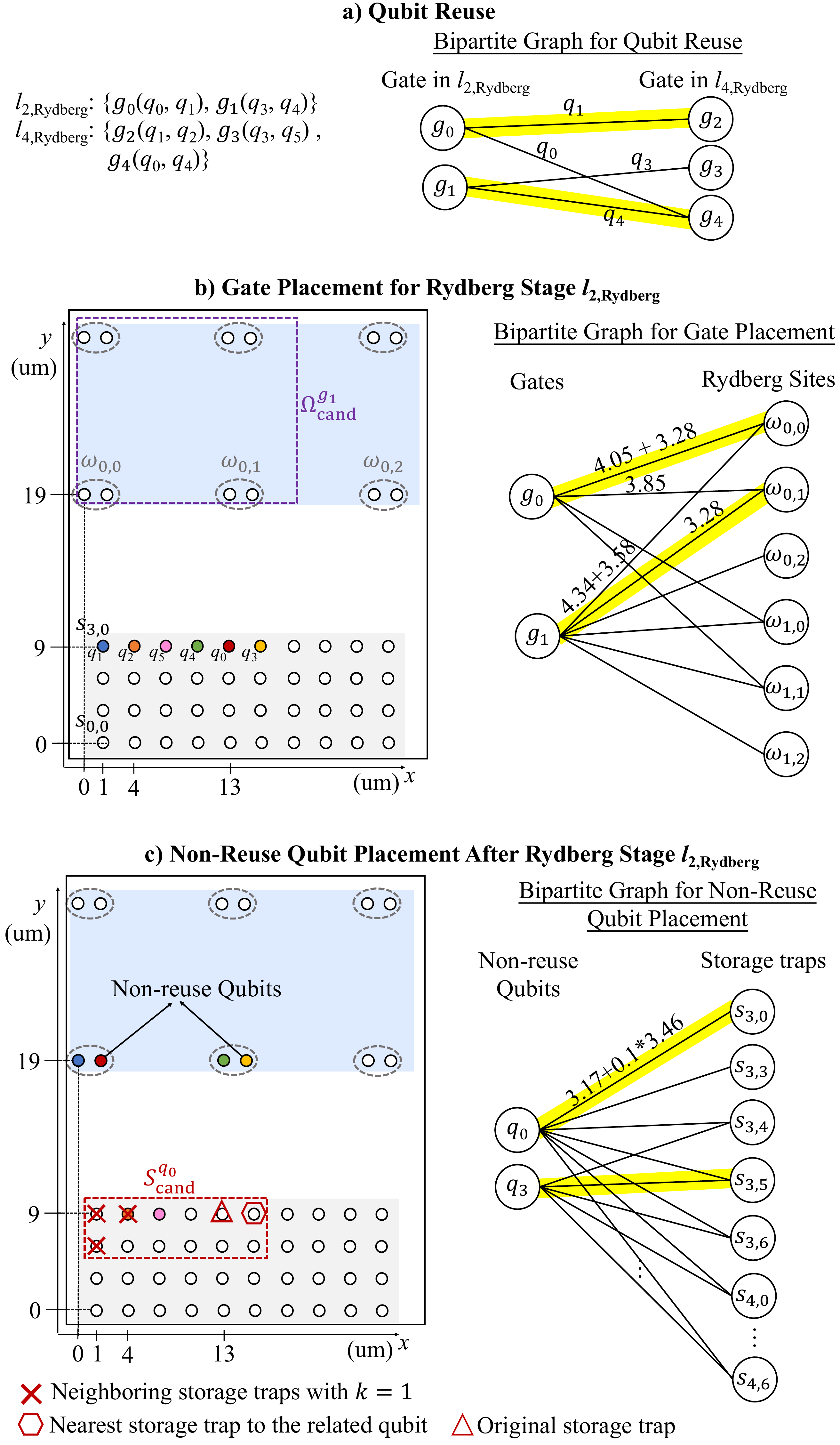}
    \caption{Intermediate placement.
    \textbf{a)} Identifying qubit reuse with bipartite matching (highlights) between gates in two Rydberg stages.
    \textbf{b)} Gate placement for Rydberg stage $l_{2,\text{Rydberg}}$.
    The qubit placement on the left is before Rydberg stage $l_{2,\text{Rydberg}}$.
    The candidate Rydberg sites $\Omega^{g_1}_\text{cand}$ are indicated by the dashed boxes.
    Gates are placed by a full matching in bipartite graph on the right.
    \textbf{c)} Non-reuse qubit placement after Rydberg stage $l_{2,\text{Rydberg}}$.
    The qubit placement on the left is based on results in b).
    The candidate storage traps for $q_0$ are $S^{q_0}_\text{cand}$, indicated by circles in dashed boxes.}
    \label{fig:intermediate_placement}
\end{figure}

\subsubsection{Qubit Reuse Strategy}
\label{subsubsec:qubit reuse}
As just mentioned, our idea is to keep reusable qubits at their current Rydberg site.
However, if both qubits at a site are reusable, keeping them both results in an error.
For instance, in Fig.~\ref{fig:intermediate_placement}a, both $q_0$ and $q_1$ in $g_0$ are reusable in the next stage: $q_0$ in $g_4$ and $q_1$ in $g_2$.
If we reuse both qubits,  $g_4$ and $g_2$ are assigned to the same site, causing a conflict.
To resolve this issue, we model the reuse relation between gates in two Rydberg stages by a bipartite graph, where vertices represent gates from each stage, and edges are drawn between vertices $v_g$ and $v_{g'}$ if a qubit of $g$ can be reused in $g'$.
To maximize the number of reuse qubits, we employ the Hopcroft–Karp algorithm~\cite{hopcroft1973n} to find the maximum cardinality matching in this bipartite graph. 
The algorithm's complexity is $O(|E|\sqrt{|V|})$. 
In our case, both vertex and edge numbers are $O(n)$, where $n$ is the number of Rydberg sites,
yielding a complexity of $O(n^{1.5})$.

\subsubsection{Gate Placement}
\label{subsubsec:gate_placement}
The qubit reuse strategy determines a set of Rydberg sites $\Omega_\text{reuse}$ occupied by gates involved with reused qubits.
We need to further place the other gates to remaining Rydberg sites while minimizing the total movement cost. 
For a placement $M$, we estimate the cost for a gate $g$ at a Rydberg site $\omega$ utilizing $\mathit{gCost}(g, \omega, M)$, as calculated in Eq.~\ref{eq:est_gate_cost}.
In addition, if gate $g(q,q')$ is reused by $g'(q,q'')$ in the next Rydberg stage, we include the movement cost of $q''$ to $\omega$, i.e., $\sqrt{d(\omega, m_{q''})}$, into the cost for gate $g$.

For this optimization, we employ a minimum weight full matching approach on a bipartite graph.
The graph vertices consist of gates and candidate Rydberg sites within $\Omega^g_\text{cand}$. 
We define $\Omega^g_\text{cand}$ as the difference between the set of neighboring sites $\Omega^g_\text{near}$ and $\Omega^g_\text{reuse}$. 
The neighboring sites are determined based on the nearest Rydberg site to the gate's target qubits with an expansion factor $\delta$ to ensure a full matching exists. 
Fig.~\ref{fig:intermediate_placement}b presents an example for $\Omega^{g_0}_\text{cand}$ and $\Omega^{g_1}_\text{cand}$.

Edges in the graph connect gates to their candidate sites, with weights representing the movement costs. 
For instance, the edge weight between $g_0$ and $\omega_{0,0}$ is calculated as 4.05+3.28 where the first term is the cost to move $q_0$ and $q_1$ to  $\omega_{0,0}$, whereas the second term is the cost for moving $q_2$ to $\omega_{0,0}$ as $q_2$ and $q_1$ will perform $g_2$ here in the next Rydberg stage.
To solve this matching problem efficiently, we utilize the Jonker-Volgenant algorithm~\cite{jonker1988shortest}, which has a time complexity of $O(|V|^3)$.
In our case, the vertices number is $O(n)$, resulting in a complexity of $O(n^3)$.
This approach allows us to optimize gate placement while considering both qubit reuse and movement minimization.

\subsubsection{Non-Reuse Dynamic Qubit Placement}
\label{subsubsec:qubit_placement}

The non-reuse qubits still needs to return to the storage zone.
Similar to gate placement, the optimal assignment of non-reuse qubits can be solved via finding the minimum-weight full matching of a bipartite graph, where vertices are the non-reuse qubits and the candidate storage traps.

For each qubit $q$, the set of candidate traps $S^q_\text{Cand}$ is constructed based on its current location and the future gate operation.
$S^q_\text{Cand}$ consists of the empty storage traps within a bounding box covering three types of the storage traps:

\begin{itemize}
    \item The qubit's original storage zone location: 
    The trap is added to ensure full matching exists because one valid solution is to return all qubits back to where they were.
    \item Neighboring storage traps close to its current Rydberg site:
    To generate these traps, we first find the nearest storage trap to the current Rydberg site, and then include its $k$-neighboring traps along the row or column.
    \item The nearest storage trap to its \textit{related qubit} (if applicable).
    We define $q'$ to be the related qubit of $q$ if $q'$ and $q$ will perform a 2Q gate in the next Rydberg stage. 
    This trap is a good candidate because placing $q$ and $q'$ near each other reduces the movement overhead for the next Rydberg stage.
    If $q$ does not involve in any gates in the next Rydberg stage, we omit this type of traps.
\end{itemize}
Fig.~\ref{fig:intermediate_placement}c provides example candidate sites for $q_0$.
First, its original trap $s_{3,4}$ is included (red triangle).
Second, the nearest storage trap to its current location $s_{3,0}$ and the $k=1$-hop storage traps to $s_{3,0}$ are included (red crosses).
Third, its related qubit is $q_4$, so the nearest storage trap to $q_4$, $s_{3,5}$ is included (red hexagon).
Lastly, we obtain the bounding box of the above traps (red dashed box), and $S_\text{cand}^{q_0}$ is all the empty traps in the bounding box.

Based on the qubit placement $M$ for a Rydberg stage, the edges connect qubit $q$ to its candidate trap $s_{r,c}$ with weights representing movement costs:
\begin{equation}
    \sqrt{\mathit{d}(s_{r,c}, m_q)}
    + \mathbf{1}_{q'}\times\alpha\times\sqrt{\mathit{d}(s_{r,c}, m_{q'})},
    \label{eq:qubit_cost}
\end{equation}
where the function $\mathbf{1}_{q'}$ indicates whether such related qubit $q'$ exists, $\alpha=0.1$ is a weighting factor for the lookahead cost.

Similar to gate placement, we apply the Jonker-Volgenant algorithm to obtain the minimum weight full matching for the qubits and traps.
In the bipartite graph, we have $O(n)$ qubit vertices, where each qubit has a constant number of candidate traps on average and $n$ is the number of Rydberg sites in entanglement zones.
Thus, the complexity for qubit placement is $O(n^3)$.
Given $O(g)$ stages in the worst case, the overall complexity for intermediate placement is $O(g \cdot n^3)$.

\section{Load-Balancing Scheduling for Multi-AOD architectures}
\label{sec:scheduling}

Based on the qubit placements, we can derive the required qubit movements.
Due to the order constraints imposed by AOD, we split the movements into multiple rearrangement jobs, such that the movements within a job can be performed by one AOD.
The detailed instructions for each rearrangement job, including atom transfer and AOD move, will be generated as described in Sec.~\ref{sec:ZAIR}.
To generate the jobs, we adopt the strategy proposed in a previous work Enola~\cite{tan2024enola}, where a maximal independent set algorithm is applied to recognize the largest set of movements that can be performed simultaneously.
According to Ref.~\cite{tan2024enola}, the complexity of generating rearrangement jobs is $O(n^2\log(n))$.

If there is only one AOD, all the rearrangement jobs are sequential. However, if there are multiple AODs, we can use them to rearrange qubits in parallel.
In scheduling, we assign the jobs to AODs and compute the exact start and end time for each instruction, including the \texttt{rydberg} instructions for 2Q gates, \texttt{1qGate} instruction for 1Q gates, and rearrangement jobs for qubit movements.
The key is to identify dependencies among the jobs.

\begin{figure}
    \centering
    \includegraphics[]{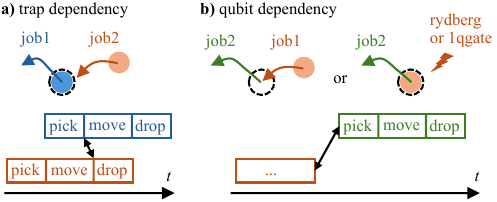}
    \caption{Two types of dependencies for rearrangement jobs.
    \textbf{a)} site dependency: job1 needs to vacate the SLM site before job2 can occupy the site.
    \textbf{b)} qubit dependency: job2 can only operate on the same qubit after job1 finishes.
    }
    \label{fig:dependency}
    \vspace{-4pt}
\end{figure}

After the machine-level instructions of each rearrangement job are derived, each job splits to three steps: picking up all the qubits, performing parallel movements of these qubits, and dropping off all the qubits.
The relative finishing time of pickup and move with respect to the beginning time of the whole job are computed.
Based on this information, we find it necessary to distinguish \textit{trap dependencies} and \textit{qubit dependencies} as illustrated in Fig.~\ref{fig:dependency}.

Trap dependency exists between two rearrangement jobs acting on the same SLM trap.
If there is a job1 moving a qubit, blue in Fig.~\ref{fig:dependency}a), currently located at an SLM trap (dashed circle), any other job that moves another qubit to this SLM trap must wait until job1 vacates the trap.
Thus, a trap dependency dictates the finishing time of the move in job2 to be later than the finishing time of the pick in job1, leaving some possibilities of overlapping between the two jobs.

Qubit dependency exists between two instructions involved with the same qubit.
For example, if two jobs move the same qubit, then there is a qubit dependency between them which forbids any time overlap between the two jobs (Fig.~\ref{fig:dependency}b).
If there is a \texttt{rydberg} or \texttt{1qGate} instruction on the qubit, the rearrangement job also needs to wait until that instruction finishes, and vice versa.

To ensure correct circuit execution, we process grouped instructions sequentially: 1) moving qubits to the Rydberg zone, 2) executing gates, 3) moving qubits back to the storage zone, and repeating this sequence until the final Rydberg stage. 
This methodology adheres to instruction dependencies.
If the instruction type is \texttt{rydberg} or \texttt{1qGate}, its start time is the latest finishing time of the instructions it depends upon.

If there are multiple AODs, we employ a load-balancing scheduling algorithm to assign parallelizable rearrangement instructions to AODs.
This algorithm iteratively allocates the longest duration instruction to the earliest available AOD.
By scheduling the longest instructions first, we minimize overall circuit execution time, preventing bottlenecks that could delay dependent instructions. 
This strategy ensures that shorter, more flexible tasks can fill any scheduling gaps, optimizing AOD utilization and maintaining high parallelism. 
Consequently, the algorithm balances the load across multiple AODs, avoiding bottlenecks and preserving the correct sequence of dependent instructions.
The start time of the arrangement instruction is determined by the later of either the available time of the assigned AOD or the finishing time of the instructions it depends on.



\section{Evaluation}
\label{sec:eval}

We implemented our proposed algorithm in Python.
We employed SciPy (v1.11.0)~\cite{2020SciPy-NMeth} for solving minimum weight full matching and maximal cardinality matching of a bipartite graph.
All experiments were conducted on an AMD EPYC 7V13 64-Core Processor at 2450 MHz and 128 GB of RAM.
Our benchmark circuits are selected from QASMBench~\cite{li2023qasmbench} with a qubit number ranging from 14 to 98 and a gate number ranging from 41 to 630.
All circuits are precompiled by Qiskit (v1.2.4) with optimization level 3~\cite{qiskit2024} for preprocessing.

\subsection{Architecture and Compiler Settings}
To demonstrate the advantage of zoned architectures, we compare with the monolithic architecture and superconducting qubit-based architectures. 
We list the architecture configurations and the compiler settings in the following paragraph:

\noindent\textbf{Zoned Architecture:} The default configuration is the architecture consisting of one AOD with 100$\times$100 sites, one storage zone with 100$\times$100 sites and one entanglement zone with 7$\times$20 sites as seen in Fig.~\ref{fig:csarchitecture}.
We compare with the leading compiler for zoned architecture, NALAC~\cite{stade2024abstract_model_for_zoned_arch}.

\noindent\textbf{Monolithic Architecture:} We consider the architecture with a single entanglement zone consisting of 10$\times$10 Rydberg sites and one AOD with 10$\times$10 sites. 
Qubit separation follows the settings of the entanglement zone in the zoned architecture. 
We compare against two state-of-the-art compilers, Atomique~\cite{wang2024atomique} and Enola~\cite{tan2024enola}.

\noindent\textbf{Superconducting Qubits:} 
We consider the IBM’s Heron superconducting machine with a 127-qubit heavy hexagon coupling graph and a 11-by-11 grid coupling graph featured Google's sycamore architecture~\cite{klimov2024snake}, and compile circuits by the default Qiskit transpiler with Sabre~\cite{li2019sabre}. 

\subsection{Fidelity Model}

\begin{table}[]
\caption{\label{tab:hardware_param} Hardware Parameters.}
    \centering
{
    \begin{tabular}{l|l|l|l|l|l}
    \hline \hline
Parameter & $f_{2}$                   & $f_1$                      & $T_\text{1q}$ & $T_\text{2q}$ & $T_2$    \\ \hline
Neutral Atom~\cite{bluvstein2024logical} & 0.995 & 0.9997 & 52us     & 360ns     & 1.5s    \\ 
SC Heron~\cite{ibmq}  & 0.999 & 0.9997 & 25ns    & 68ns     & 311us \\
SC Grid~\cite{klimov2024snake}  & 0.999 & 0.9997 & 25ns    & 42ns     & 89us \\\hline \hline
\end{tabular}}
\end{table}

We consider three main error sources: imperfect gates, atom transfers, and qubit decoherence. 
We assume that the qubit movement affects circuit fidelity by increasing decoherence errors.
Given a movement distance $d$, we calculate the movement time $t$ based on the relation $d/t^2=2750$m/s$^2$~\cite{bluvstein2022quantum}.
According to the experiments in~\cite{bluvstein2022quantum}, the qubit movement at this speed does not incur fidelity decrease or atom loss.

The hardware parameters are derived from the leading hardware~\cite{bluvstein2022quantum,bluvstein2024logical}.
To be conservative, we assume 1Q gates are executed sequentially with the fidelity $f_1=99.97\%$ and duration $T_\text{1q}=52$us.
This estimation is based on conservative pulse duration~\cite{bluvstein2024logical}: three Rz and two Ry pulses, each 8us, and the separation between two pulses is 3us.
CZ gates are implemented by a global Rydberg laser with the fidelity $f_2=99.5\%$ and duration $T_\text{Ryd}=360$ns.
Idle qubits excited by the Rydberg laser results in the fidelity $f_\text{exc}=99.75\%$.
Multiple atoms can be transferred simultaneously from one tweezer to the other in $T_\text{tran}=15$us,
and each transfer has the fidelity $f_\text{tran}=99.9\%$, accounting for both dephasing and atom loss.

The decoherence error is estimated by a linear model $1-t_q/T_2$, where $t_q$ is the idling time for qubit $q$ and $T_2=1.5$s is the coherence time for neutral atoms.
The linear model is a reasonable approximation in scenarios where $t_q\ll T_2$.
The qubit idling time is the time that a qubit is not performing gates or atom transfers.

The total circuit fidelity is computed by
\begin{equation*}
    f=\underbrace{(f_{1})^{g_1}}_\text{1Q gate}\cdot \underbrace{(f_{2})^{g_2} \cdot (f_\text{exc})^{N_\text{exc}}}_\text{2Q gate} \cdot \underbrace{(f_\text{tran})^{N_\text{tran}}}_\text{atom transfer} \cdot \underbrace{\Pi_{q\in Q}\ (1-\frac{t_q}{T_2})}_\text{decoherence}, \label{eq:fidelity}
\end{equation*}
where $g_1$ and $g_2$ are the number of 1Q and 2Q gates, $N_\text{exc}$ is the number of qubits excited by the Rydberg laser without performing a 2Q gate, $N_\text{tran}$ is the number of atom transfers, and $Q$ is the set of qubits.

The hardware parameters for superconducting qubits are from the leading commercial machines.
For the Heron architecture, we extract the parameters of ibm\_torino from IBMQ platform~\cite{mckay2023benchmarking_ibm,ibmq}.
We set $T_2$ = 311us,  and use 2Q gate fidelity $f_2$ = 99.9$\%$~\cite{mckay2023benchmarking_ibm}.
When estimating the decoherence error, we use $68$ns for 2Q gate duration and $25$ns for 1Q gate duration~\cite{ibmq}.
The parameters for the grid architecture are derived from~\cite{klimov2024snake}, with $T_2$ = 89us and $T_\text{2q}$ = 42ns.
The parameters are summarized in~Table~\ref{tab:hardware_param}.

\begin{figure*}[t]
    \centering
    \includegraphics[width=\linewidth]{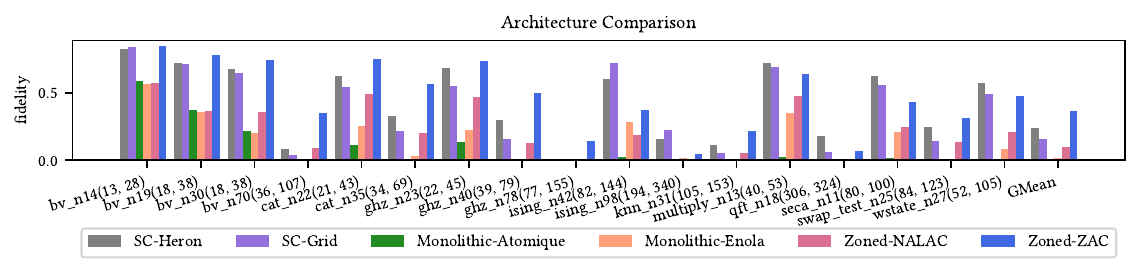}
    \vspace{-20pt}
    \caption{Circuit fidelity comparisons across different architectures. 
    `SC' means superconducting qubits.
    The name of the benchmark circuit is followed by the number of program qubits and the gate number. For example, qft\_n18(306/324) represents the QFT circuit with 18 qubits, 306 2Q gates and 324 1Q gates.}
    \label{fig:architecture_comparison}
    \vspace{-4pt}
\end{figure*}

\subsection{Monolithic Architecture vs Zoned Architecture vs Superconducting Qubit}
In our evaluation results, the performance gain is summarized by the geometric mean.
Fig.~\ref{fig:architecture_comparison} demonstrates that ZAC outperforms all baselines for every circuit.
\textit{We achieve a 1.56$\times$ and 2.33$\times$ fidelity improvement compared to the superconducting Heron and grid architectures, respectively,
due to better decoherence fidelity.}
Table~\ref{table:sc_zac} shows the fidelity breakdown and average circuit duration for the Heron architecture and ZAC,
where each fidelity term is the geometric mean among all benchmarks.
Although the zoned architecture based on neutral atom have longer circuit duration compared to the superconducting qubit platform due to slow operations and qubit movement, 
it has less decoherence errors due to the long coherence time.
However, with better 2Q gate fidelity, for circuits with short circuit duration, superconducting qubit architectures may have advantages over the zoned architecture, 
e.g., ising\_n42 has a fidelity of 0.601 and 0.722 with duration 2us and 650ns on SC Heron and Grid, respectively,  with while its fidelity on the zoned architecture is 0.37 with duration 10,439us.

\begin{figure*}[t]
    \centering
    \includegraphics[width=\linewidth]{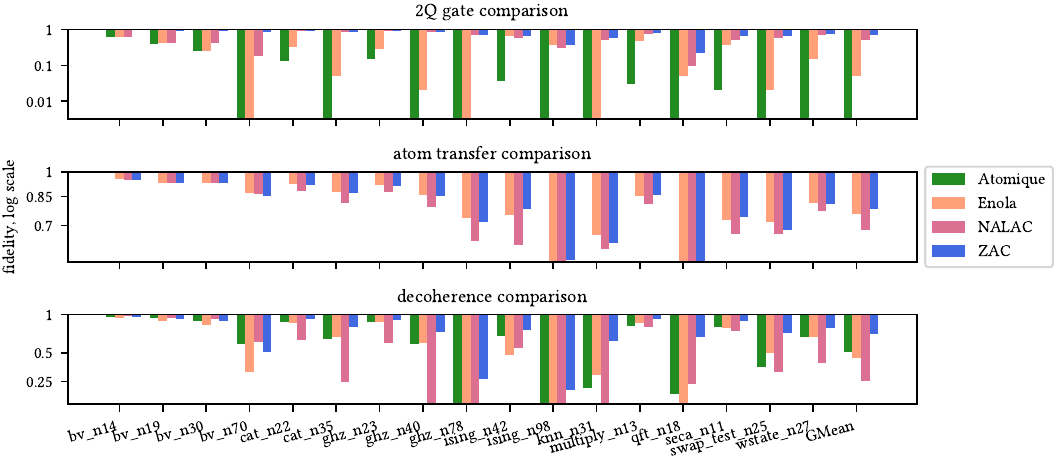}
    \caption{Fidelity breakdown for Atomique, Enola, NALAC and ZAC.
    Shorter bars are better because of the log scale.}
    \label{fig:fidelity_breakdown}
\end{figure*}

\begin{figure*}[t]
    \centering
    \includegraphics[width=\linewidth]{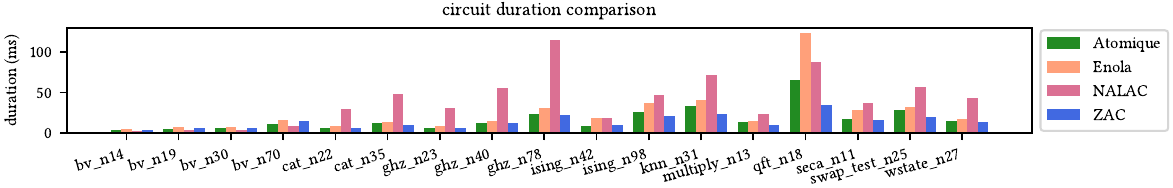}
    \caption{Circuit duration for Atomique, Enola, NALAC and ZAC.}
    \label{fig:cir_duration}
\end{figure*}

\color{black}
\textit{Compared with the monolithic architecture, ZAC increases fidelity by 22$\times$ and 13,350$\times$ compared to Enola and Atomique, respectively.}
Moreover, for zoned architectures, ZAC delivers a 4$\times$ fidelity improvement over NALAC.
Note that in terms of 2Q gate fidelity, NALAC has addition qubit excitation errors since they keep idle qubits in the entanglement zone to enhance qubit reuse.
Fig.~\ref{fig:fidelity_breakdown} provides the fidelity breakdown for ZAC, NALAC, Enola, and Atomnique.
Without qubit excitation errors caused by the Rydberg laser, ZAC achieves 1.37$\times$ and 14$\times$ better 2Q gate fidelity than NALAC and Enola.
In terms of atom transfer fidelity, Atomique does not utilize atom transfers but employs SWAP gates to change qubit locations.
Compared with Enola, ZAC demonstrates a 1.03$\times$ improvement in atom transfer fidelity via exploiting qubit reuse.
For decoherence errors, ZAC exhibits a 1.36$\times$ fidelity improvement compared with Atomique.
This improvement can be attributed to the shorter qubit traveling distance.
Fig.~\ref{fig:cir_duration} illustrates the circuit duration.
Since qubits can be placed more compactly in the storage zone, 
we have shorter movement distances to bring qubits to the Rydberg site compared to the monolithic architecture.
However, NALAC has the longer circuit duration for large cases, showing that their placement strategy and reuse strategy fail to reduce the movement overhead.
Overall, ZAC achieves 10\% and 55\% shorter circuit duration compared to Atomique and NALAC, respectively.

We see performance differences across various architectures depending on the circuit characteristics.
For circuits with low parallelism such as BV (Bernstein-Vazirani), GHZ (Greenberger-Horne-Zeilinger), QFT (Quantum Fourier Transform), where the gate execution is sequential in nature, 
the zoned architecture demonstrates larger fidelity improvement compared to the monolithic architecture.
Because the qubit excitation errors for the monolithic architecture is detrimental, the circuit performance drops rapidly as the qubit number grows.
For example, for bv\_n70, ZAC shows an 635$\times$ fidelity improvement compared with the monolithic architecture.

On the other hand, for circuits with high parallelism like Ising, the ability to handle multiple qubit interactions simultaneously is critical. 
In such cases, monolithic architectures, which can be viewed to maximize qubit reuse, may display a stronger performance as they are inherently designed to support dense and simultaneous qubit interactions without the overhead of frequent zone transfers. 
However, zoned architectures, when optimized, can still be competitive by leveraging qubit reuse and strategic placement to minimize movement overhead.
Furthermore, the compact layout in the storage zones further reduces decoherence errors, enhancing the overall performance.
Therefore, for circuits with high parallelism such as ising\_n98, ZAC can still deliver a 11$\times$ fidelity gain. 
In conclusion, while zoned architectures excel in maintaining high fidelity for sequential circuits by minimizing qubit excitation errors, they also exhibit competitive performance for highly parallel circuits through minimizing movement overhead and decoherence errors.

\begin{table}[t]
\caption{\label{table:sc_zac} Fidelity breakdown and average circuit duration for the superconducting qubit grid architecture and ZAC.
}
{\begin{tabular}{l|l|l|l|l|l|l}
\hline \hline
& \multicolumn{5}{c|}{Fidelity breakdown} & Avg. \\ \cline{2-6}
& 2Q gate & 1Q gate &  Tran. & Decohe. & Total & duration \\ \hline
SC  & 0.8451  & 0.9008  & N/A  & 0.3102    &  0.2362 & 9.1us \\
ZAC   & 0.6977  & 0.9721  & 0.7814  & 0.7003  & 0.3689 & 13.8ms \\ \hline \hline        
\end{tabular}}
\end{table}

\subsection{Ablation Study of Compilation Techniques}
\begin{figure*}[t]
    \centering
    \includegraphics[width=\linewidth]{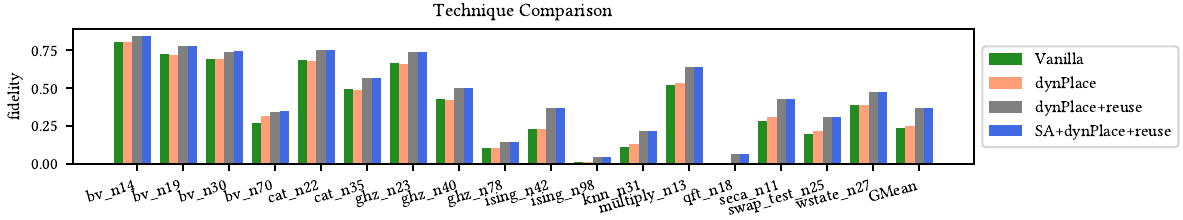}
    \caption{Circuit fidelity comparison for different settings in ZAC.
    `Vanilla' means trivial initial placement and fix intermediate qubit placement without any qubit reuse. 
    `dynPlace' means dynamic qubit placement. 
    `reuse' means reuse-aware qubit placement.
    `SA' means SA-based initial placement.
    }
    \label{fig:abalation-study}
\end{figure*}

Fig.~\ref{fig:abalation-study} presents a comparative analysis of different techniques in ZAC.
`Vanilla' is the baseline approach with trivial and static qubit placement, where we place qubits sequentially according to their indices, 
starting from the first storage trap in the nearest row to the entanglement zone.
By introducing dynamic placement, `dynPlace' achieves a 5\% fidelity improvement, 
showing its effectiveness in reducing the average distance of qubit movements.
`dynPlace+reuse' further incorporates qubit reuse and boosts the fidelity by 46\% compared to `dynPlace'.
As the reuse of qubits minimizes unnecessary movements, it lessens the errors associated with atom transfer and decoherence.
`SA+dynPlace+reuse' attains the best performance by adding simulated annealing-based initial placement.
This setting only demonstrates a 0.4\% fidelity increase on average.
Since the qubits in the circuits fit into one row of the storage zone, all qubits have the same distance to the entanglement zone, and the effect of initial placement is less significant.
However, it realizes up to a 4\% gain on circuits such as qft\_n18.
This evaluation results highlight the effectiveness of integrating advanced placement strategies and reuse policies in enhancing quantum circuit fidelity.


\begin{figure}[t]
    \centering\includegraphics[width=0.9\linewidth]{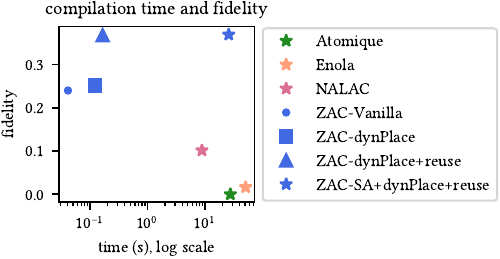}
    \caption{
    Comparison of the average compilation time and circuit fidelity in the geometric mean across all circuits for Atomique, Enola, NALAC, and ZAC.
    Each color represents a compiler, 
    and different marker indicates the strategy adopted in ZAC.}
    \label{fig:runtime}
    \vspace{-1em}
\end{figure}

\subsection{Scalablity Study}
Fig.~\ref{fig:runtime} reveals the trade-off between fidelity and the compilation time. 
ZAC demonstrates its efficiency by achieving higher fidelity with runtime comparable to other tools. 
If we disable the optimization for initial mapping, ZAC can solve every instances in less than 1 second, achieving 63$\times$ speedup and 3.6$\times$ better fidelity compared to NALAC.
The results show that ZAC is a more effective and efficient compiler compared to other tools.

\subsection{Optimality Study}
\begin{figure*}[t]
    \centering
    \includegraphics[width=\linewidth]{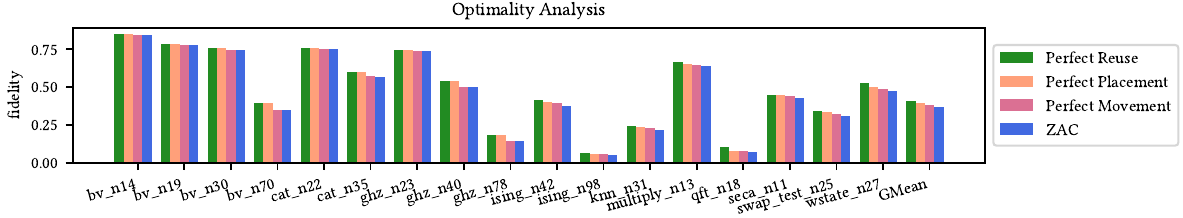}
    \vspace{-15pt}
    \caption{Optimality analysis.
    `Perfect movement' considers the ideal case when all qubit movement can be performed in parallel.
    `Perfect placement' assumes the shortest movement between zones.
    `Perfect reuse' considers maximum qubit reuse via direct movements between Rydberg sites.}
    \label{fig:optimality_study}
    \vspace{-5pt}
\end{figure*}

In this experiment, we evaluate the optimality gap for ZAC by comparing it against the fidelity of three ideal cases, perfect reuse, perfect movement and perfect placement.
Note that perfect placement is built upon the assumption of perfect movement, 
while perfect reuse incorporates both perfect placement and perfect movement. 
These idealized scenarios represent upper bounds on fidelity, though they are not generally achievable. 
The fidelity represents the overall circuit fidelity.
The evaluation results are illustrated in Fig.~\ref{fig:optimality_study}.

The setting for perfect movement assumes that all movements are compatible.
Therefore, between two Rydberg instructions, we have at most two rearrangement instructions:
the first rearrangement instruction moves the qubits that are not involved in gates from the entanglement zone to the storage zone,
and the second rearrangement instruction relocates
the qubits, that are about to go through gates, from the storage zone to the entanglement zone.
ZAC only demonstrates a 3\% optimality gap compared with the perfect movement case, showing that our placement strategy effectively maximizes the movement parallelism.

Based on the assumptions of ideal movement, perfect placement assumes that the distance to move a qubit between a storage trap and a Rydberg site is the zone separation. 
Thus, the duration for each rearrangement layer is the minimum possible duration for any rearrangement layer, i.e., $2T_\text{tran}+T_\text{move}=2T_\text{tran}+\sqrt{d_\text{sep}/a}$.
Under this setting, ZAC has a 7\% optimality gap, indicating that our placement strategy effectively minimizes the movement duration.

The setting for perfect reuse represents the most ideal scenario for zoned architectures. 
With the assumptions in perfect placement, we further assume that for the reusable qubit can stay at the current site or be directly moved to its next site, reducing two atom transfers required for moving the qubit back to a storage trap.
\textit{Compared with perfect reuse, ZAC only demonstrates a 10\% optimally gap.
The optimality study reveals that our compiler's performance is near optimal, demonstrating its effectiveness for zoned architectures.}

\color{black}

\subsection{Effectiveness of Multiple AODs}
\begin{figure*}[t]
    \centering
    \includegraphics[width=\linewidth]{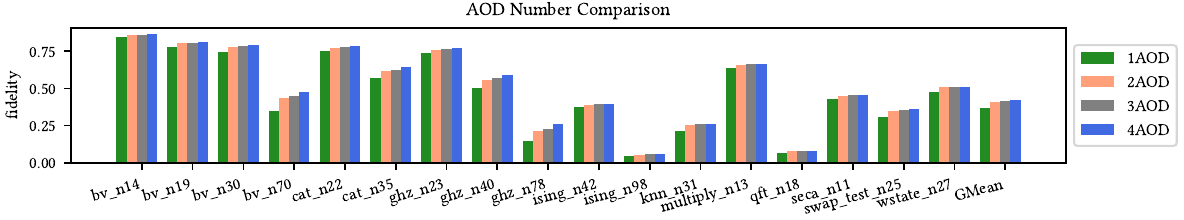}
    \vspace{-15pt}
    \caption{Architecture evaluation with different AOD numbers.}
    \vspace{-5pt}
    \label{fig:comp-aod}
\end{figure*}

Fig.~\ref{fig:comp-aod} demonstrates the impact of utilizing multiple AODs on circuit fidelity across various benchmarks.
\textit{The use of multiple AODs generally enhances circuit fidelity, with the greatest improvements observed in having two AODs with a 10\% fidelity improvement.}
This increase comes from the increased parallelism for rearrangement instructions.
The improvement becomes less significant with more than two AODs since there are not enough rearrangement instructions to fully utilize the increased parallelism.
Thus, adding the third and fourth AOD only gives a 2\% fidelity improvement.
The gain of multiple AODs observed in Atomique can be more significant than ZAC 
because Atomique reduces movement distance and Rydberg stages in a monolithic architecture via multiple AODs.
However, in zoned architectures, the movement distance is fixed, and the number of Rydberg stages does not affect fidelity, so the gain is more moderate.

\subsection{Effectiveness of Multiple Entanglement Zones}
\label{subsec:eval_mez}
We demonstrate the flexibility of ZAC by evaluating the circuit fidelity on different zone layouts, e.g., multiple entanglement zones.
The advantage of having multiple entanglement zones is to reduce the average distance between a storage trap and a Rydberg site.
The zone size of the default configuration is too large for our circuits to demonstrate the advantage of the second entanglement zone.
Because we only utilize one row in the storage zone and two to three rows in the entanglement zone,
adding the second entanglement zone does not reduce the movement distance.
Therefore, we consider a smaller architecture where qubits occupy multiple rows in the storage zone.
Thus, when adding the second entanglement zone, we can effectively reduce the movement distance.

\begin{figure}
    \centering
    \includegraphics[width=\linewidth]{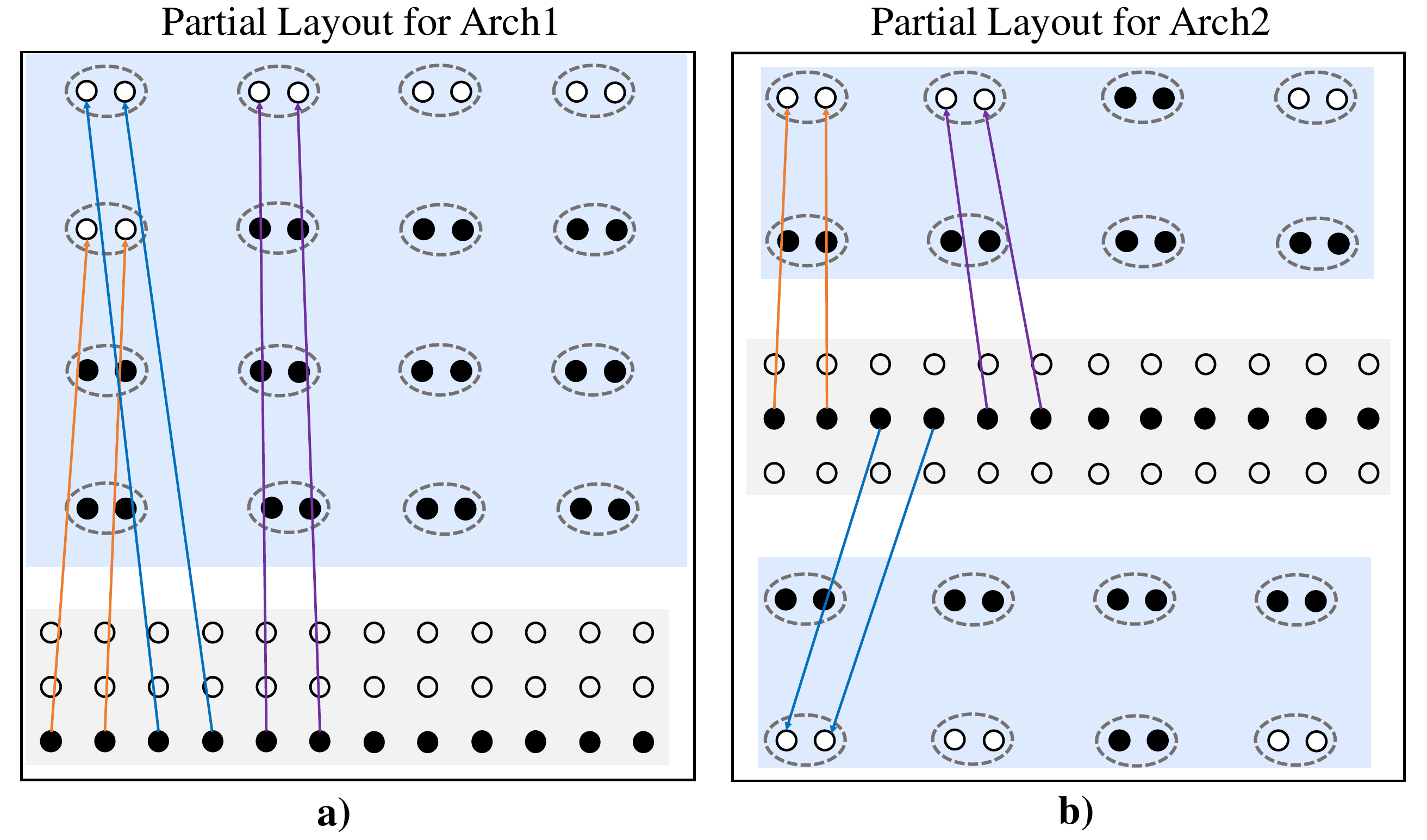}
    \caption{Demonstration for qubit movement of circuit ising\_n98.
    \textbf{a)} Partial architecture for Arch1.
    \textbf{b)} Partial architecture for Arch2.}
    \label{fig:ising_n98}
    \vspace{-4pt}
\end{figure}

In this experiment, we consider circuit ising\_n98 with 98 qubits.
To have a fair comparison, both architectures have the same configuration for the storage zone and the number of Rydberg sites.
Arch1 is the small architecture with 3$\times$40 storage traps, and an entanglement zone with 6$\times$10 sites.
Arch2 is the architecture with two entanglement zones, each comprising of 3$\times$10 sites located above and below the storage zone as indicated in Fig.~\ref{fig:ising_n98}.
Ising\_n98 is a highly parallel circuit, where 49 2Q gates can be executed simultaneously.
Therefore, the qubits fill up most of the entanglement sites. 
With a single entanglement zone, qubits need to travel long distance to reach the sites at the rear row as shown in Fig.~\ref{fig:ising_n98}a.
The performance for ising\_n98 on Arch1 has a circuit fidelity of 0.041 and a circuit duration of 23.25ms.
With the second entanglement zone, the fidelity on Arch2 is 0.047, which is a 15\% improvement
because when equipped with two entanglement zones, the distance to reach the rear sites are reduced~as illustrates in Fig.~\ref{fig:ising_n98}b.
The circuit duration is also shortened to 21.63ms, which is a 8\% reduction.
This result demonstrates the potential of architectures with multiple entanglement zones to improve circuit performance in terms of both fidelity and efficiency.
Additionally, the experiment demonstrates that ZAC is capable of supporting architectures with multiple entanglement zones. 


\section{ZAC in Fault-Tolerant Quantum Computing}
\label{sec:ftqc}

\begin{figure}[t]
    \centering
    \includegraphics[width=\linewidth]{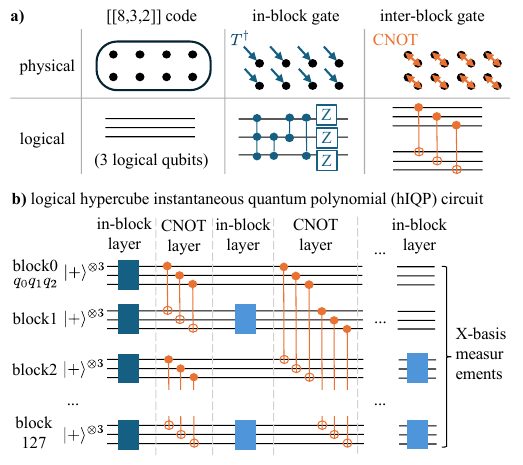}
    \caption{
    \textbf{a)} An [[8,3,2]] code block encodes 3 logical qubits with 8 physical qubits. 
    The physical layout is 2 rows by 4 columns.
    An in-block logical gate equivalent to CCZ, CZ, and $Z$ gates can be realized by applying physical $T^\dag$ gates on all qubits.
    Inter-block CNOT gates can be realized by (transversal) CNOTs on corresponding qubits of two blocks.
    \textbf{b)} A hypercube instantaneous quantum polynomial (hIQP) circuit on 128 [[8,3,2]] code blocks (384 logical qubits).
    This is a scaled-up version of the hIQP circuit in \cite{bluvstein2024logical}.
    All logical qubits are initiated in $|+\rangle$ and measured in the $X$ basis.
    There are 8 in-block gate layers interleaved with 7 (inter-block) CNOT layers.
    The stride of CNOTs doubles each time:
    in the first layer, the CNOTs are on (0,1), (2,3), ..., (126,127); in the second layer, they are on (0,2), (1,3), ..., (125,127), etc.
    }
     \vspace{-6pt}
    \label{fig:iqp}
\end{figure}

Recent experiments in reconfigurable neutral atom arrays have demonstrated fault-tolerant quantum algorithms using tens of logical qubits~\cite{bluvstein2024logical}, highlighting the need for compilation support for fault-tolerant quantum computing (FTQC).

In FTQC, logical qubits are redundantly encoded into more physical qubits to enable the detection and correction of quantum errors.
For instance, the [[8,3,2]] code, as summarized in Fig.~\ref{fig:iqp}a, encodes 3 logical qubits using 8 physical qubits with distance 2, allowing detection of any single-qubit error~\cite{vasmer2022832code}.
To implement logical gates, physical gates are carefully designed to manipulate logical qubits while remaining within the code space.
Transversal gates are broadly defined as gates that do not propagate errors within the same code block, making them preferred in FTQC.
In addition, reconfigurable architectures such as trapped ions and neutral atoms can directly perform transversal gates leveraging available long-range connectivity.
Two example transversal gates of the [[8,3,2]] code are displayed in Fig.~\ref{fig:iqp}a: the `in-block gate,' where applying physical $T^\dag$ gates equates to a combination of logical CCZ, CZ, and $Z$ gates; and the `inter-block gate,' where applying physical CNOTs to corresponding qubits in two blocks equates to logical CNOTs on corresponding logical qubits.

The [[8,3,2]] codes were recently used to implement hypercube instantaneous quantum polynomial (hIQP) circuits~\cite{bluvstein2024logical, compilingIQP}.
This class of circuits can be implemented using transversal gates and may offer quantum advantage, as sampling from these circuits could solve certain \#P-hard counting problems~\cite{bremner2016iqp}.
We consider an hIQP circuit with 384 logical qubits encoded in 128 [[8,3,2]] code blocks and 448 transversal gates.
In this circuit, there are layers of in-block gates interleaved with inter-block CNOT layers.
The stride of the CNOTs increases by 2 in each layer, generating a hypercube connectivity, e.g., in the first layer, CNOTs connect blocks 0 and 1, 2 and 3, etc.; in the second layer, CNOTs connect blocks 0 and 2, 1 and 3, and so on.
This circuit is a scaled-up version of the 48-qubit circuit from Ref.~\cite{bluvstein2024logical}, likely beyond the reach of state-of-the-art classical simulation methods~\cite{maslov2024fastclassicalsimulationharvardquera}.

In the FTQC context, ZAC serves two main purposes.
First, it can compile physical qubit movements for FTQC subroutines such as syndrome extraction and logical gates.
For the hIQP circuits on [[8,3,2]] codes, these tasks are straightforward, so we omit them here.
Second, by inputting the logical circuit consisting of transversal gates, ZAC can determine the movements of logical code blocks to implement the correct CNOTs between them.
The physical qubits within each logical block move together.
Thus, ZAC’s qubit movement strategy directly informs how logical blocks should be repositioned to execute transversal gates.
In the hIQP example, we apply ZAC to the 128 code blocks and a smaller logical-level architecture.
Since according to the settings in~\cite{bluvstein2024logical}, each code block occupies 2 rows by 4 columns, and the physical architecture includes 7 rows and 20 columns of sites in the entanglement zone, the logical-level architecture supports $\lfloor7/2\rfloor=3$ rows and $\lfloor20/4\rfloor=5$ columns in the entanglement zone.
Ref.~\cite{bluvstein2024logical} provides a movement heuristic that extends to cases where both the number of rows and columns are powers of 2, using only $2\times4=8$ sites out of the 15 available.
In contrast, ZAC can leverage all available sites, resulting in a compiled output with 35 Rydberg stages and a physical circuit duration of 117.847 ms.

The results above demonstrate that ZAC is an effective tool for the movements required by transversal CNOTs between code blocks.  
However, this addresses only one of the aspects in FTQC compilation for neutral atom arrays, and further work is needed to advance toward large-scale universal computation.

\color{black}
\section{ZAIR: an Intermediate Representation for Zoned Architectures}
\label{sec:ZAIR}

\begin{figure}
    \centering
    \includegraphics[width=\linewidth]{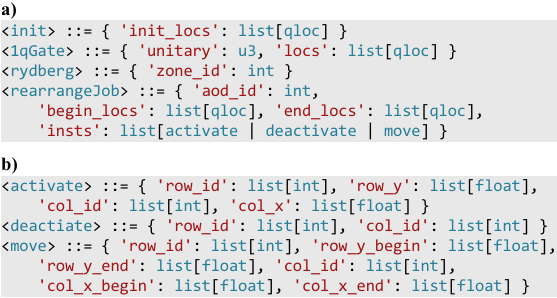}
    \caption{
    \textbf{a)} Four types of instructions in ZAIR.
    \textbf{b)} Three types of machine-level instructions involved in a \texttt{rearrangeJob} instruction in ZAIR.}
    \label{fig:ir}
    \vspace{-6pt}
\end{figure}

Given a quantum circuit, the ultimate output of our compiler are machine-level instructions on the controllable components including AODs, the Rydberg laser (for 2Q gates), and the Raman lasers (for 1Q gates).
However, just like on classical computers, the number of machine-level instructions can grow quickly, hiding the structure of program execution.
To tackle with this complexity, we introduce ZAIR (\underline{z}oned \underline{a}rchitecture \underline{i}ntermediate \underline{r}epresentation) as a level of abstraction.
ZAIR has four types of instructions listed in Fig.~\ref{fig:ir}a.
It locates a qubit in SLM with \texttt{qloc} that is a 4-tuple $(q, a, r, c)$ meaning qubit $q$ is at row $r$ and column $c$ of SLM array $a$.

The \texttt{init} instruction appears only once in the beginning for the initial location of each qubit.
\texttt{1qGate} means applying the 1Q gate \texttt{unitary} to a set of \texttt{qloc}s.
\texttt{rydberg} means applying the Rydberg laser to a specific entanglement zone with index \texttt{zone\_id}.
Each of the \texttt{rydberg} instruction executes what we call a \textit{Rydberg stage} which is a set of 2Q gates applied in parallel by the same Rydberg laser exposure.
A Rydberg stage requires qubit pairs to be properly moved together to an entanglement zone to apply the 2Q gates.

The three instructions discussed above are at the machine-level.
In contrast, the \texttt{rearrangeJob} instruction corresponds to a rearrangement job, which will be broken down to multiple machine-level instructions.
The \texttt{begin\_locs} is comprised of a list of the \texttt{qloc}s for qubits that will be rearranged by the AOD.
The \texttt{end\_locs} has the same shape as \texttt{begin\_locs} but the \texttt{qloc}s here change to the ending locations after the rearrangement.
For example, in Fig.~\ref{fig:csarchitecture}, qubits 0-3 form a little ``square'' at the bottom left of the storage zone, i.e., they occupy (0,0), (0,1), (1,0), and (1,1) in SLM array 0. 
Thus, \texttt{begin\_locs}=[[(0,0,0,0), (1,0,0,1)], [(2,0,1,0), (3,0,1,1)]].
The indicated rearrangement job (red dashes) moves them to the entanglement zone with \texttt{end\_locs} =[[(0,1,0,2), (1,2,0,2)], [(2,1,1,2), (3,2,1,2)]].
$q_0$ and $q_1$ are at Rydberg site $\omega_{0,2}$ whereas $q_2$ and $q_3$ are at $\omega_{1,2}$.

\begin{figure}
    \centering
    \includegraphics[width=\linewidth]{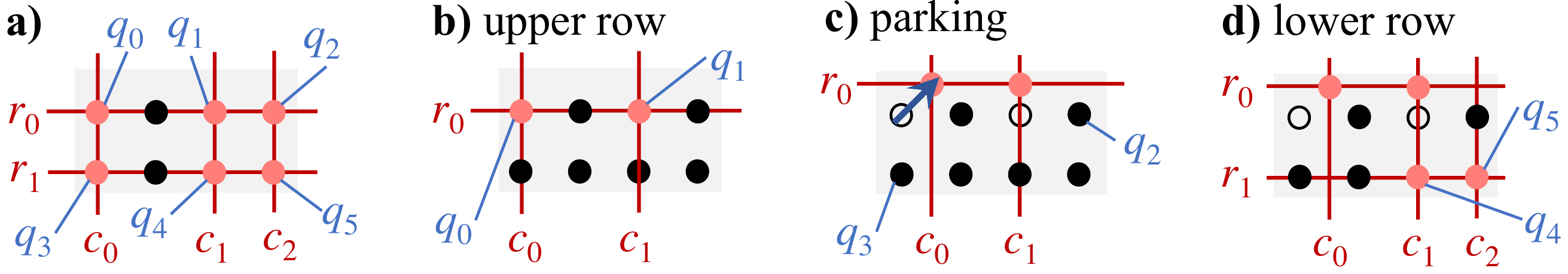}
    \caption{
    \textbf{a)} When we turn on the three AOD columns and two AOD rows, all $q_0$ to $q_5$ are picked up.
    However, suppose we do not want to pick up $q_2$ and $q_3$. 
    We can do the following.
    \textbf{b)} Activate $r_0$, $c_0$ and $c_1$ so that $q_0$ and $q_1$ are picked up at the intersections.
    \textbf{c)} Perform a parking movement (arrow) that shifts $r_0$ and $c_0$ so that $q_2$ does not align with $r_0$, and $q_3$ does not align with $c_0$.
    \textbf{d)} Activate $r_1$ and $c_2$ so that $q_4$ and $q_5$ are picked up at the intersections.
    }
    \label{fig:parking}
    \vspace{-8pt}
\end{figure}

The \texttt{insts} in a rearrangement job is populated by machine-level AOD instructions listed in Fig.~\ref{fig:ir}b.
There are some details to consider in generating these machine-level instructions.
For example, other than the movements that shuttle qubits between zones, we may need some small movements dubbed \textit{parking} during the pickup phase in a job.
When we activate a set of AOD rows and columns, all qubits at their intersections will be picked up as exhibited in Fig.~\ref{fig:parking}a.
However, the rearrangement job may not be performing on a set of qubits with this structure, the so-called combinatorial rectangle~\cite{tan_depth-optimal_2024}, e.g., we may not want to move $q_2$ and $q_3$.
In this case, we can operate in a few steps as shown in Fig.~\ref{fig:parking}b--d.
In this work, we adopt a simple pickup strategy following OLSQ-DPQA~\cite{tan2024dpqa} where we turn on the AOD row by row and possibly insert parking between activating two rows.

The rearrangement job in ZAIR is at a sufficient level of abstraction for the purpose of our compilation since it serves as a natural interface between the placement in Sec.~\ref{sec:placement} and the scheduling in Sec.~\ref{sec:scheduling}.
Generating machine-level instructions from rearrangement jobs can involve specific machine details, so this may be done by the hardware provider instead of the user. 
No matter how a job is instantiated to machine-level instructions, it occupies a worker (AOD) for a continuous period of time, so our definition makes it convenient to leverage multiple AODs.
During the scheduling in the compilation, each job is assigned to an AOD by setting the \texttt{aod\_id} field.
ZAIR is also efficient: among the benchmark set in this paper, the number of ZAIR instructions per gate is 0.85 geomean; the number of machine-level instructions per gate is 1.77 in geometry mean.
The number of ZAIR instruction can be lower than the number of gates because many 2Q gates can be implemented together in a few ZAIR instructions in benchmarks with high parallelism like Ising.

\color{black}

\section{Conclusion}
\label{sec:conclusion}
In this work, we propose ZAC, a compiler for zoned quantum architectures based on neutral atoms.
Our compiler demonstrates significant improvement compared to with the monolithic architecture or the superconducting qubit platforms.
By leveraging qubit mobility and innovative placement and scheduling techniques, 
we address the critical challenges associated with Rydberg excitation errors and qubit movement overhead. 
Our proposed method effectively mitigates idle qubit excitation errors by isolating qubits in storage zones when not in use, enhancing overall circuit fidelity. 
Additionally, reuse-aware placement strategies minimize the total movement between zones, further reducing errors and improving performance.
Lastly, our load-balancing scheduling algorithm ensures efficient distribution of rearrangement instructions, maximizing parallelism and minimizing execution time.
The experimental results highlight the effectiveness of our compiler, showing that it achieves near-optimal performance with only a 10\% gap from the ideal solution. 
Compared with the leading compiler for zoned architectures, we demonstrate a 4$\times$ fidelity improvement.
The flexibility of our approach is demonstrated through 
its adaptability to advanced architectures with multiple AODs and multiple entanglement zones.

We also propose a comprehensive specification for zoned architectures, supporting the expression with multiple entanglement and storage zones.
In addition, we introduce the corresponding intermediate representation, ZAIR, that abstracts rearrangement operations into jobs. 
This abstraction facilitates the use of multiple AODs to move qubits in parallel.

Overall, this work underscores the potential of zoned architectures in realizing scalable and reliable quantum computing with neutral atoms, 
and the ability to support FTQC makes ZAC a versatile tool for both current quantum computations and future FTQC applications.
There are several future research directions. 
First, to further improve circuit fidelity, we may explore other optimizations such as allowing movements within entanglement zones for more advanced qubit reuse.
Second, our compiler can be extended to support circuits with mid-circuit readout and multi-qubit gates by revising the cost function.
Lastly, our evaluation result for multiple entanglement zones demonstrates the circuit performance can be improved by running on architectures with different configurations.
Similar concepts have been proposed for superconducting qubit platforms~\cite{lin2022domain,yang2023superconducting,liang2023superconducting,yang2024processor}.
This result suggests the possibilities for tailoring quantum processors to diverse computational tasks, 
further enhancing circuit efficiency and performance.

\section*{Acknowledgements}
This work is partially funded by NSF grants CCF-2313083 and OSI-2410716.
The authors thank Y. Stade, M. Cain, D. Bluvstein, P. Liu, H. Zhou, M. Kalinowski, and Prof. M. D. Lukin for valuable discussions.

\bibliographystyle{IEEEtranS}
\bibliography{refs}


%
%
%
%
%




\newcommand{\zacurl}{\url{https://doi.org/10.5281/zenodo.14219335}}


\appendix
\section{Artifact Appendix}

\subsection{Abstract}

The artifact includes the ZAC compiler, baseline compilers for comparison, benchmarks, and all necessary scripts to run experiments and generate figures.
It provides instructions for using the ZAC framework to replicate key results, including: (1) performance comparisons presented in Fig.~\ref{fig:architecture_comparison}, Fig.~\ref{fig:fidelity_breakdown}, Fig.~\ref{fig:cir_duration},  Fig.~\ref{fig:abalation-study}, and Table~\ref{table:sc_zac}
and (2) architecture evaluation shown in Fig.~\ref{fig:comp-aod} and Section~\ref{subsec:eval_mez}.
The process is divided into two main steps: (1) data generation and (2) figure plotting. 
For brevity, instructions for remaining experiments are omitted.

\subsection{Artifact check-list (meta-information)}
{\small
\begin{itemize}
  \item {\bf Algorithm:} Simulated Annealing, Hopcroft–Karp Algorithm, Jonker-Volgenant Algorithm, Maximal Independent Set Algorithm
  \item {\bf Program:} Python
  \item {\bf Compilation:} Python 3.10 with Qiskit 1.2.4 and Scipy. All are public available.
  \item {\bf Run-time environment: }Python with Qiskit. A Linux system is recommended.
  \item {\bf Hardware: } A CPU with 8 cores and 16 GB of memory is recommended.
  \item {\bf Execution: }No specific condition is required.
  \item {\bf Metrics: } Quantum circuit duration and circuit fidelity
  \item {\bf Output: } Figures 8-11 and 14 of the article and the necessary data for Table~\ref{table:sc_zac} and  Section~\ref{subsec:eval_mez}.
  \item {\bf Experiments: } Please see Section~\ref{subsec:exp_flow} for detailed instructions for reproducing the results
  \item {\bf How much disk space required (approximately)?: } $<$10G
  \item {\bf How much time is needed to prepare workflow (approximately)?: } $<$10 minutes
  \item {\bf How much time is needed to complete experiments (approximately)?: } $<$2 hours
  \item {\bf Publicly available?: } Our compiler is available at \zacurl
  \item {\bf Code licenses (if publicly available)?:} BSD 3-Clause License
  \item {\bf Archived (provide DOI)?: } \zacurl
\end{itemize}
}

\subsection{Description}

\subsubsection{How to access}

The artifact is available at the following link: \zacurl

\subsubsection{Hardware dependencies}
To complete the experiments in a reasonable amount of time, a CPU with 8 cores and 16 GB of memory is recommended.

\subsubsection{Software dependencies}
The artifact is implemented in Python and requires several packages including Qiskit. 
A complete list of required packages can be found in the requirements.txt in the artifact.

\subsection{Installation}

For artifact evaluation, first download the artifact from Zenodo at \zacurl and unzip the file.
Then, you can install the required python packages using pip with following commands: 

\texttt{\$ cd ZAC\_AE}

\texttt{\$ pip install -r requirements.txt}

\subsection{Experiment workflow}
\label{subsec:exp_flow}

\subsubsection{Data generation}
Once the required packages are installed, one can generate data via the command

\texttt{\$ ./gen\_results.sh}

The script contains four steps.
The first step is to generate simulation results for superconducting qubit platforms, including the grid architecture and heavy-hexagon architecture. 
Next, we proceed to generate the simulation results for the monolithic neutral atom architecture in the second and third step using the baseline compiler, Enola and Atomique, respectively.
Lastly, we generate the results of ZAC, including ablation study, AOD number comparison, and zone configuration comparison.
The produced log files are in directory ``log". 
A line ``[INFO] Finish Compilation" should appear in the end of the log if the compiler are executed successfully.

Baseline results for the NALAC compiler are included in the artifact. 
Due to NALAC's complicated installation process and nondeterministic behavior, its execution is omitted from the main workflow. 
The source code for NALAC is provided in the artifact under the directory ``other\_compiler/nalac\_source".
Users can follow the README in the directory to execute NALAC.

\subsubsection{Figure Generation}
After collecting results from all compilers,  run to plot figure.

\texttt{\$ python process\_data.py}

\subsection{Evaluation and expected results}

The compiled circuits produced by the compiler are organized into a directory structure under the root directory ``result". 
Each subdirectory consists of a set of compiled circuits from a compiler and contains the following folders:
\begin{itemize}
    \item code: Circuit instructions in the ZAIR format.
    \item fidelity: Simulated fidelity of the circuit.
    \item time: Logs the runtime of the solver used during compilation.
\end{itemize}

After the figures are generated, they are stored in the ``fig" directory for visualization, and the csv files for plotting the figures are stored in the ``fig" directory for analysis.

\subsection{Experiment customization}
Users can test ZAC on their own benchmarks and customize experiments by configuring the compilation settings according to the instructions provided in the README.md file included with the artifact.

\subsection{ZAIR Example}
\begin{figure}[t]
    \centering
    \includegraphics[width=\linewidth]{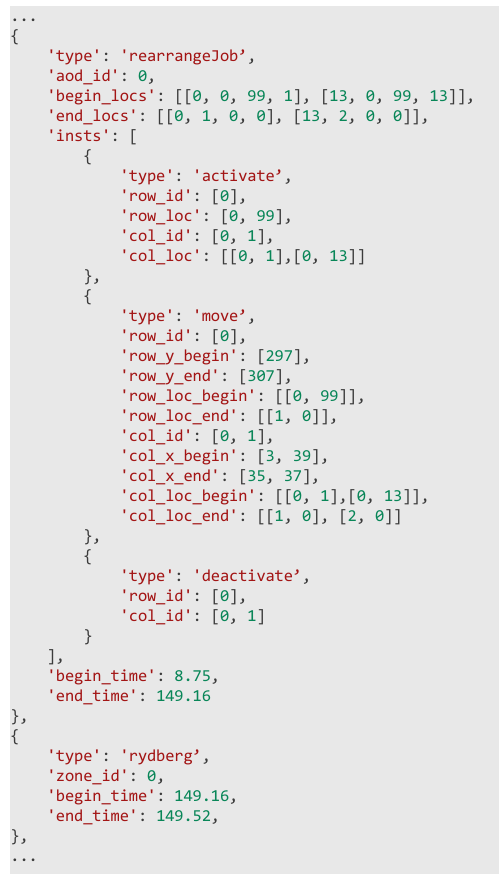}
    \caption{A \texttt{rearrangeJob} and \texttt{Rydberg} instruction in ZAIR from the compiled bv\_n14 circuit.}
    \label{fig:ir_example_bv}
\end{figure}

\begin{figure}[t]
    \centering
    \includegraphics[width=\linewidth]{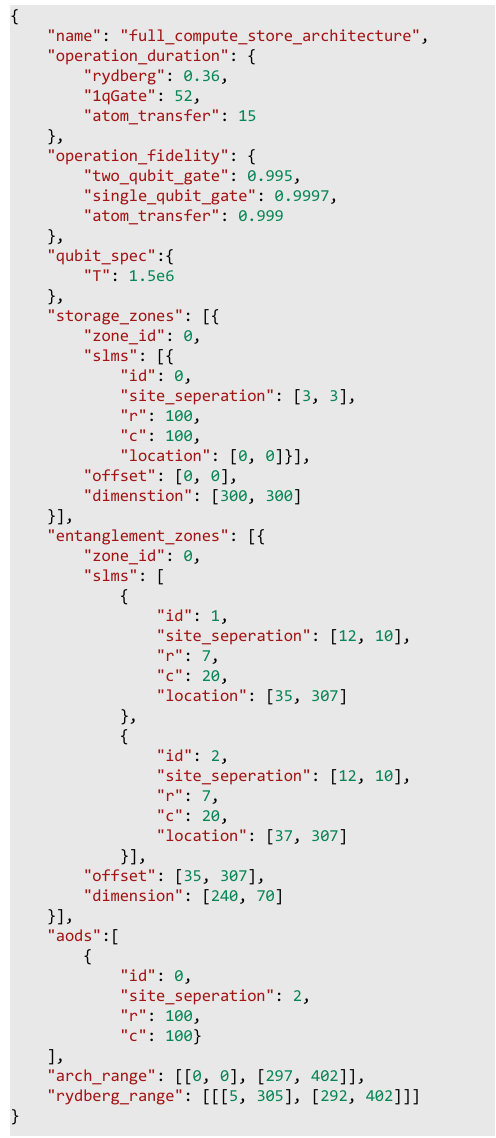}
    \caption{Specification of the reference zoned architecture.}
    \label{fig:arch_example}
\end{figure}

In this section, we provide an example for ZAIR.
Fig.~\ref{fig:ir_example_bv} shows two instructions for implementing the first CZ gate of the 14-qubit circuit bv\_n14, implemented the Bernstein–Vazirani algorithm, on the architecture in Fig.~\ref{fig:csarchitecture}.
The specification of this architecture is provided in Fig.~\ref{fig:arch_example}.

Initially, all qubits are placed according to their indices in the storage zone, i.e., $q_i$ is located at trap $(0, 99, i)$.
A \texttt{rearrangeJob} instruction performed by AOD-0 moves $q_0$ and $q_{13}$ from traps  $(0,99,1)$ and $(0,99,13)$ in the storage zone to traps $(1,0,0)$ and $(2,0,0)$ in the entanglement zone, respectively.
First, we activate row 0 and columns 0 and 1 of AOD-0 at row 99 and columns 0 and 13 in SLM-0 to transfer $q_0$ and $q_{13}$ from SLM-0 to AOD-0.
After moving qubits to their target locations in the entanglement zone, we deactivate row and columns of AOD-0 to transfer the qubits to SLM-1 and SLM-2.
Lastly, a Rydberg laser is turned on to perform the CZ gate.

\subsection{Methodology}

Submission, reviewing and badging methodology:

\begin{itemize}
  \item \url{https://www.acm.org/publications/policies/artifact-review-and-badging-current}
  \item \url{https://cTuning.org/ae}
\end{itemize}

\makeatletter
\setlength{\@fptop}{0pt}
\makeatother

\end{document}